\documentclass[
	amsmath,
	amssymb,
	a4paper,
	aps,		
	prx,		
	reprint,	
	fleqn,
	showpacs,
	longbibliography,
	floatfix
]{revtex4-1}

\usepackage[T1]{fontenc}
\usepackage{lmodern}
\usepackage{microtype}

\usepackage{tabularx}
\usepackage{booktabs}
\usepackage{multirow}

\usepackage{graphicx}
\usepackage{dcolumn}
\usepackage{bm}
\usepackage{subfig}
\usepackage[version=3]{mhchem}

\usepackage{amssymb}
\usepackage{amsthm}
\usepackage{color}
\usepackage{gensymb}
\usepackage{wasysym}
\usepackage[
	font=small,
	format=hang,
	indention=-1.0cm,
	justification=RaggedRight
]{caption}

\newcommand{\myTitle}{Nonequilibrium Thermodynamics of Chemical Reaction Networks: \\ Wisdom from Stochastic Thermodynamics}

\usepackage{hyperref}

\newcommand{\myName}{Riccardo Rao}
\newcommand{\myAffiliation}{Complex Systems and Statistical Mechanics, Physics and Materials Science Research Unit, University of Luxembourg, L-1511 Luxembourg, Luxembourg}

\newcommand{\myAdvisor}{Massimiliano Esposito}

\newcommand{\derpart}[2]{\frac{\partial{#1}}{\partial{#2}}}

\newcommand{\de}{\mathrm{d}}
\newcommand{\dt}{\mathrm{d}_{t}}

\newcommand{\ave}[1]{\left\langle {#1} \right\rangle}
\newcommand{\abs}[1]{\left\lvert{#1}\right\rvert}
\newcommand{\Order}[1]{\mathrm{O}\left({#1}\right)}

\newcommand{\transpose}{^{\mathrm{T}}}

\newcommand{\kb}{k_\mathrm{B}}

\newcommand{\epr}{\dot{S}_{\mathrm{i}}}
\newcommand{\efr}{\dot{S}_{\mathrm{e}}}
\newcommand{\cwr}{\dot{W}_{\mathrm{c}}}
\newcommand{\dwr}{\dot{W}_{\mathrm{d}}}
\newcommand{\vepr}[1]{\dot{S}_{\mathrm{#1}}}

\newcommand{\sset}[1]{\left\{ #1 \right\}}
\newcommand{\set}[1]{\{ #1 \}}

\newcommand{\dr}[1]{\Delta_{\mathrm{r}}#1_{\rho}}

\newcommand{\drs}[1]{\Delta_{\mathrm{r}}#1_{\rho}^{\circ}}

\newcommand{\Ly}[3]{#3{\mathcal{L}}\left( \{ #1^{\sigma} \} | \{ #2^{\sigma} \} \right)}
\newcommand{\LyX}[2]{\mathcal{L}\left( \{ #1^{\sigma_{\mathrm{x}}} \} | \{ #2^{\sigma_{\mathrm{x}}} \} \right)}

  {\left\lbrace\begin{array}{@{}l@{}}}%
  {\end{array}\right.}

\theoremstyle{definition} 
\newtheorem{example}{Example}

\theoremstyle{plain}

\definecolor{butter1}{rgb}{0.98,0.91,0.31}
\definecolor{butter2}{rgb}{0.93,0.83,0}
\definecolor{butter3}{rgb}{0.77,0.63,0}
\definecolor{skyblue1}{rgb}{0.45,0.62,0.81}
\definecolor{skyblue2}{rgb}{0.2,0.39,0.64}
\definecolor{skyblue3}{rgb}{0.13,0.29,0.53}
\definecolor{scarlet1}{rgb}{0.93,0.16,0.16}
\definecolor{scarlet2}{rgb}{0.8,0,0}
\definecolor{scarlet3}{rgb}{0.64,0,0}
\definecolor{chameleon1}{rgb}{0.54,0.88,0.2}
\definecolor{chameleon2}{rgb}{0.45,0.82,0.09}
\definecolor{chameleon3}{rgb}{0.3,0.6,0.02}
\definecolor{orange1}{rgb}{0.98,0.68,0.24}
\definecolor{orange2}{rgb}{0.96,0.47,0}
\definecolor{orange3}{rgb}{0.8,0.36,0}
\definecolor{plum1}{rgb}{0.68,0.5,0.66}
\definecolor{plum2}{rgb}{0.46,0.31,0.48}
\definecolor{plum3}{rgb}{0.36,0.21,0.4}
\definecolor{chocolate1}{rgb}{0.91,0.72,0.43}
\definecolor{chocolate2}{rgb}{0.75,0.49,0.07}
\definecolor{chocolate3}{rgb}{0.56,0.35,0.01}
\definecolor{aluminium1}{rgb}{0.93,0.93,0.92}
\definecolor{aluminium2}{rgb}{0.82,0.84,0.81}
\definecolor{aluminium3}{rgb}{0.73,0.74,0.71}
\definecolor{aluminium4}{rgb}{0.53,0.54,0.52}
\definecolor{aluminium5}{rgb}{0.33,0.34,0.32}
\definecolor{aluminium6}{rgb}{0.18,0.2,0.21}  

\definecolor{webgreen}{rgb}{0,.5,0}
\definecolor{webbrown}{rgb}{.6,0,0}
\definecolor{grigio}{rgb}{.85,.85,.85} 
\definecolor{RoyalBlue}{rgb}{0.0, 0.14, 0.4}

\hypersetup{%
    colorlinks=true, linktocpage=true, pdfstartpage=1, pdfstartview=FitV,%
    breaklinks=true, pdfpagemode=UseNone, pageanchor=true, pdfpagemode=UseOutlines,%
    plainpages=false, bookmarksnumbered, bookmarksopen=true, bookmarksopenlevel=1,%
    hypertexnames=true, pdfhighlight=/O,
    urlcolor=webbrown, linkcolor=RoyalBlue, citecolor=webgreen, 
    pdftitle={\myTitle},%
    pdfauthor={\textcopyright\ \myName},%
    pdfsubject={},%
    pdfkeywords={},%
    pdfcreator={pdfLaTeX},%
    pdfproducer={LaTeX REVTeX}%
}

\begin{document}


\title{\myTitle}

\author{\myName}
\affiliation{\myAffiliation}
\author{\myAdvisor}
\affiliation{\myAffiliation}

\date{\today. Published in \emph{Phys.~Rev.~X}, DOI:~\href{https://doi.org/10.1103/PhysRevX.6.041064}{10.1103/PhysRevX.6.041064}}

\begin{abstract}
	We build a rigorous nonequilibrium thermodynamic description for open chemical reaction networks of elementary reactions (CRNs).
	Their dynamics is described by deterministic rate equations with mass action kinetics.
	Our most general framework considers open networks driven by time-dependent chemostats.
	The energy and entropy balances are established and a nonequilibrium Gibbs free energy is introduced.
	The difference between this latter and its equilibrium form represents the minimal work done by the chemostats to bring the network to its nonequilibrium state.
	It is minimized in nondriven detailed-balanced networks (i.e. networks which relax to equilibrium states) and has an interesting information-theoretic interpretation.
	We further show that the entropy production of complex balanced networks (i.e. networks which relax to special kinds of nonequilibrium steady states) splits into two non-negative contributions:
	one characterizing the dissipation of the nonequilibrium steady state and the other the transients due to relaxation and driving.
	Our theory lays the path to study time-dependent energy and information transduction in biochemical networks.
\end{abstract}

\pacs{
05.70.Ln,  
87.16.Yc   
}

\maketitle

\section{Introduction}

Thermodynamics of chemical reactions has a long history. 
The second half of the 19th century witnessed the dawn of the modern studies on thermodynamics of chemical mixtures.
It is indeed at that time that J.~W.~Gibbs introduced the concept of \emph{chemical potential} and used it to define the thermodynamic potentials of non-interacting mixtures \cite{gibbs61}.
Several decades later, this enabled T.~de~Donder to approach the study of chemical reacting mixtures from a thermodynamic standpoint.
He proposed the concept of \emph{affinity} to characterize the chemical force irreversibly driving chemical reactions and related it to the thermodynamic properties of mixtures established by Gibbs \cite{donder27}.
I.~Prigogine, who perpetuated the Brussels School founded by de~Donder, introduced the assumption of \emph{local equilibrium} to describe irreversible processes in terms of equilibrium quantities \cite{prigogine47,prigogine67}.
In doing so, he pioneered the connections between thermodynamics and kinetics of chemical reacting mixtures \cite{prigogine54}.

During the second half of the 20th century, part of the attention moved to systems with small particle numbers which are ill-described by ``deterministic'' rate equations.
The Brussels School, as well as other groups, produced various studies on the nonequilibrium thermodynamics of chemical systems \cite{oster73,malek-mansour75,schnakenberg76,hill77,mou86,ross08}
using a stochastic description based on the (Chemical) Master Equation \cite{mcquarrie67,gillespie92}.
These studies played an important role during the first decade of the 21st century for the development of \emph{Stochastic Thermodynamics}, a theory that systematically establishes a nonequilibrium thermodynamic description for systems obeying stochastic dynamics \cite{sekimoto10,jarzynski11,seifert12,vandenbroeck15}, including chemical reaction networks (CRNs) \cite{gaspard04,andrieux04,schmiedl07:biomolecules,schmiedl07,polettini15}.

Another significant part of the attention moved to the thermodynamic description of biochemical reactions in terms of deterministic rate equations \cite{alberty03,beard08}.
This is not so surprising since living systems are the paramount example of nonequilibrium processes and they are powered by chemical reactions. 
The fact that metabolic processes involve thousands of coupled reactions also emphasized the importance of a network description \cite{palsson06,palsson11,palsson15}.
While complex dynamical behaviors such as oscillations were analyzed in small CRNs \cite{goldbeter96,epstein98}, most studies on large biochemical networks focused on the steady-state dynamics.  
Very few studies considered the thermodynamic properties of CRNs \cite{beard02,qian03,beard04,chakrabarti13}.
One of the first nonequilibrium thermodynamic description of open biochemical networks was proposed in Ref.~\cite{qian05}.
However, it did not take advantage of Chemical Reaction Network Theory which connects the network topology to its dynamical behavior and which was extensively studied by mathematicians during the seventies \cite{horn72,feinberg72,horn72:complex} (this theory was also later extended to stochastic dynamics \cite{kurtz72,anderson10,anderson15,cappelletti16}).
As far as we know, the first and single study that related the nonequilibrium thermodynamics of CRNs to their topology is Ref.~\cite{polettini15}, still restricting itself to steady states.

In this paper, we consider the most general setting for the study of CRNs, namely open networks driven by chemostatted concentrations which may change over time.
To the best of our knowledge, this was never considered before.
In this way, steady-state properties as well as transient ones are captured.
Hence, in the same way that Stochastic Thermodynamics is built on top of stochastic dynamics, we systematically build a nonequilibrium thermodynamic description of CRNs on top of deterministic chemical rate equations.
In doing so, we establish the energy and entropy balance and introduce the nonequilibrium entropy of the CRN as well as its nonequilibrium Gibbs free energy.
We show this latter to bear an information-theoretical interpretation similar to that of Stochastic Thermodynamics \cite{takara10,esposito11,sagawa13,parrondo15} and to be related to the dynamical potentials derived by mathematicians.
We also show the relation between the minimal chemical work necessary to manipulate the CRNs far from equilibrium and the nonequilibrium Gibbs free energy.
Our theory embeds both the Prigoginian approach to thermodynamics of irreversible processes \cite{prigogine54} and the thermodynamics of biochemical reactions \cite{alberty03}.
Making full use of the mathematical Chemical Reaction Network Theory, we further analyze the thermodynamic behavior of two important classes of CRNs: detailed-balanced networks and complex-balanced networks.
In absence of time-dependent driving, the former converges to thermodynamic equilibrium by minimizing their nonequilibrium Gibbs free energy.
In contrast, the latter converges to a specific class of nonequilibrium steady states and always allow for an adiabatic--nonadiabatic separation of their entropy production, which is analogous to that found in Stochastic Thermodynamics \cite{esposito07,harris07,esposito10:threefaces1,vandenbroeck10,ge10}. 
We note that while finalizing this paper, a result similar to the latter was independently found in Ref. \cite{ge16}.

\subsection*{Outline and Notation}

The paper is organized as follows.
After introducing the necessary concepts in chemical kinetics and Chemical Reaction Network Theory, sec.~\ref{sec:crn}, the nonequilibrium thermodynamic description is established, sec.~\ref{sec:thermodynamics}.
As in Stochastic Thermodynamics, we build it on top of the dynamics and formulate the entropy and energy balance, \S~\ref{sec:entropyBalance} and \S~\ref{sec:energyBalance}. Chemical work and nonequilibrium Gibbs free energy are also defined and the information-theoretic content of the latter is discussed.
The special properties of detailed-balance and of complex-balanced networks are considered in sec.~\ref{sec:dbnThermo} and \ref{sec:cbnThermo}, respectively.
Conclusions and perspectives are drawn in sec.~\ref{sec:conclusions}, while some technical derivations are detailed in the appendices.

We now proceed by fixing the notation.
We consider a system composed of reacting \emph{chemical species} $\ce{X}_{\sigma}$, each of which is identified by an index $\sigma \in \mathcal{S}$, where $\mathcal{S}$ is the set of all indices/species.
The species populations change due to \emph{elementary reactions}, \emph{i.e.} all reacting species and reactions must be resolved (none can be hidden), 
and all reactions must be \emph{reversible}, \emph{i.e.} each forward reaction $+\rho$ has a corresponding backward reaction $-\rho$.
Each pair of forward--backward reactions is a \emph{reaction pathway} denoted by $\rho \in \mathcal{R}$.
The orientation of the set of reaction pathways $\mathcal{R}$ is arbitrary.
Hence, a generic CRN is represented as
\begin{equation}
	\sum_{\sigma} \nabla^{\sigma}_{+ \rho} \, \ce{X}_{\sigma} 
	\overset{k^{+\rho}}{\underset{k^{-\rho}}{\rightleftharpoons}} 
	\sum_{\sigma} \nabla^{\sigma}_{- \rho} \, \ce{X}_{\sigma} \, .
	\label{reaction}
\end{equation}
The constants $k^{+\rho}$ ($k^{-\rho}$) are the \emph{rate constants} of the forward (backward) reactions.
The stoichiometric coefficients $-\nabla^{\sigma}_{+\rho}$ and $\nabla^{\sigma}_{-\rho}$ identify the number of molecules of $\ce{X}_{\sigma}$ involved in each forward reaction $+\rho$ (the stoichiometric coefficients of the backward reactions have opposite signs).
Once stacked into two non-negative matrices $\nabla_{+} = \set{\nabla^{\sigma}_{+\rho}}$ and $\nabla_{-} = \set{\nabla^{\sigma}_{-\rho}}$, they define the integer-valued \emph{stoichiometric matrix}
\begin{equation}
	\nabla \equiv \nabla_{-} - \nabla_{+} \, .
	\label{eq:sm}
\end{equation}
The reason for the choice of the symbol ``$\nabla$'' will become clear later.

\begin{figure}[t]
	\centering
	\includegraphics[width=.48\textwidth]{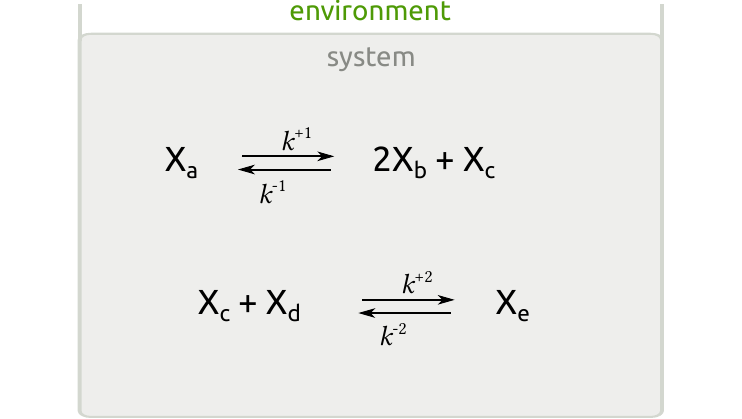}
	\caption{
		Representation of a closed CRN.
		The chemical species are $\set{\ce{X}_{\mathrm{a}},\cdots,\ce{X}_{\mathrm{e}}}$.
		The two reaction pathways are labeled by $1$ and $2$.
		The nonzero stoichiometric coefficients are $- \nabla^{\mathrm{a}}_{+1} = - 1$, $\nabla^{\mathrm{b}}_{-1} = 2$ and $\nabla^{\mathrm{c}}_{-1} = 1$ for the first forward reaction and $- \nabla^{\mathrm{c}}_{+2} = - 1$, $- \nabla^{\mathrm{d}}_{+2} = - 1$ and $\nabla^{\mathrm{e}}_{-2} = 1$ for the second one.
		Since the network is closed, no chemical species is exchanged with the environment.
	}
	\label{fig:cCRN}
\end{figure}

\begin{example}
	The stoichiometric matrix of the CRN depicted in fig.~\ref{fig:cCRN} is
	\begin{equation}
		\nabla =
		\begin{pmatrix}
			-1	& 0 \\
			2	& 0 \\
			1	& -1 \\
			0	& -1 \\
			0	& 1 
		\end{pmatrix} \, .
		\label{ex:smExample}
	\end{equation}
	\qed
\end{example}

Physical quantities associated to species and reactions are represented in upper--lower indices vectorial notation.
Upper and lower indexed quantities have the same physical values, \emph{e.g.} $Z^{i} = Z_{i}, \forall \, i$.
We use the Einstein summation notation: repeated upper--lower indices implies the summation over all the allowed values for those indices---\emph{e.g.} $\sigma \in \mathcal{S}$ for species and $\rho \in \mathcal{R}$ for reactions.
Given two arbitrary vectorial quantities $\mathbf{a} = \set{a^{i}}$ and $\mathbf{b} = \set{b^{i}}$, the following notation will be used
\begin{equation*}
	{a^{i}}^{b_{i}} \equiv \prod_{i} {a^{i}}^{b_{i}} \, . 
\end{equation*}
Finally, given the matrix ${C}$, whose elements are $\set{C^{i}_{j}}$, the elements of the transposed matrix ${C}\transpose$ are $\set{C^{j}_{i}}$.

The time derivative of a physical quantity $A$ is denoted by $\dt{A}$, its steady state value by an overbar, $\bar{A}$, and its equilibrium value by $A_{\mathrm{eq}}$ or $A^{\mathrm{eq}}$.
We reserve the overdot, $\dot{A}$, to denote the rate of change of quantities which are \emph{not} exact time derivatives.

\section{Dynamics of CRNs}
\label{sec:crn}

In this section, we formulate the mathematical description of CRNs \cite{feinberg:lectures,gunawardena03} in a suitable way for a thermodynamic analysis.
We introduce closed and open CRNs and show how to drive these latter in a time-dependent way.
We then define conservation laws and cycles and review the dynamical properties of two important classes of CRNs: detailed-balanced networks and complex-balanced networks.

We consider a chemical system in which the reacting species $\set{\ce{X}_{\sigma}}$ are part of a homogeneous and ideal dilute solution:
the reactions proceed slowly compared to diffusion and the solvent is much more abundant than the reacting species.
Temperature $T$ and pressure $p$ are kept constant. 
Since the volume of the solution $V$ is overwhelmingly dominated by the solvent, it is assumed constant.
The species abundances are large enough so that the molecules discreteness can be neglected.
Thus, at any time $t$, the system state is well-described by the molar \emph{concentration distribution} $\set{ Z^{\sigma} \equiv {N^{\sigma}}/{V}}$, where $N^{\sigma}$ is the molarity of the species $\ce{X}_{\sigma}$.

The reaction kinetics is controlled by the \emph{reaction rate functions} $J^{\pm \rho} \big(\sset{ Z^{\sigma} } \big)$, which measure the rate of occurrence of reactions and satisfy the \emph{mass action kinetics} \cite{groot84,feinberg:lectures,pekar05}
\begin{equation}
J^{\pm \rho} \equiv J^{\pm \rho} \big( \sset{ Z^{\sigma} } \big) = k^{\pm \rho} {Z^{\sigma}}^{\nabla^{\pm \rho}_{\sigma}} \, .
\label{ex:mak}
\end{equation}
The net \emph{concentration current} along a reaction pathway $\rho$ is thus given by
\begin{equation}
		J^{\rho} \equiv J^{+\rho} - J^{-\rho} 
		= k^{+ \rho} {Z^{\sigma}}^{\nabla^{+ \rho}_{\sigma}} - k^{- \rho} {Z^{\sigma}}^{\nabla^{- \rho}_{\sigma}} \, .
	\label{ex:currents}
\end{equation}

\begin{example}
	For the CRN in fig.~\ref{fig:cCRN} the currents are
	\begin{equation}
		\begin{aligned}
			J^{1} & = k^{+1} Z^{\mathrm{a}} - k^{-1} (Z^{\mathrm{b}})^{2} Z^{\mathrm{c}} \\
			J^{2} & = k^{+2} Z^{\mathrm{c}} Z^{\mathrm{d}} - k^{-2} Z^{\mathrm{e}} \, .
		\end{aligned}
		\label{}
	\end{equation}
	\qed
\end{example}

\subsection{Closed CRNs}
\label{sec:closed}

A closed CRN does not exchange any chemical species with the environment.
Hence, the species concentrations vary solely due to chemical reactions and satisfy the rate equations
\begin{equation}
	\dt{Z}^\sigma = \nabla^\sigma_{\rho} \, J^{\rho} , \quad \forall \, \sigma \in \mathcal{S} \, .
	\label{eq:rate}
\end{equation}
Since rate equations are nonlinear, complex dynamical behaviors may emerge \cite{epstein98}.
The fact that the rate equations \eqref{eq:rate} can be thought of as a \emph{continuity equation} for the concentration, where the stoichiometric matrix $\nabla$ \eqref{eq:sm} acts as a discrete differential operator, explains the choice of the symbol ``$\nabla$'' for the stoichiometric matrix \cite{polettini14}.

\subsection{Driven CRNs}
\label{sec:drivencrn}

\begin{figure}[t]
	\centering
	\includegraphics[width=.48\textwidth]{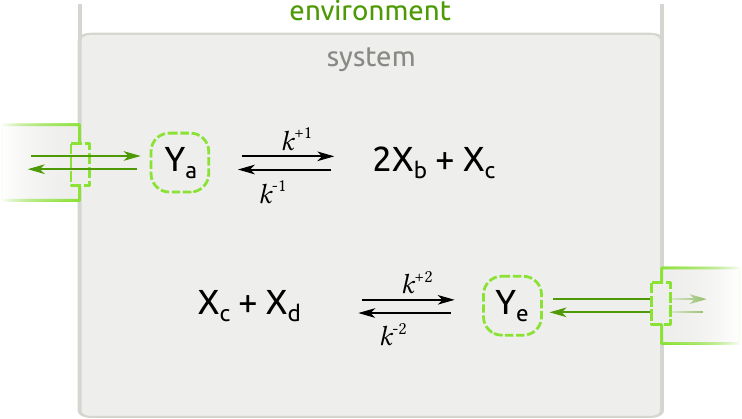}
	\caption{
		Representation of an open CRN.
		With respect to the CRN in fig.~\ref{fig:cCRN}, the species $\ce{X}_{\mathrm{a}}$ and $\ce{X}_{\mathrm{e}}$ are chemostatted, hence represented as $\ce{Y}_{\mathrm{a}}$ and $\ce{Y}_{\mathrm{e}}$.
		The green boxes aside represent the reservoirs of chemostatted species.
	}
	\label{fig:oCRN}
\end{figure}

In open CRNs, matter is exchanged with the environment via reservoirs which control the concentrations of some specific species, fig.~\ref{fig:oCRN}.
These externally controlled species are said to be \emph{chemostatted}, while the reservoirs controlling them are called \emph{chemostats}.
The chemostatting procedure may mimic various types of controls by the environment.
	For instance, a direct control could be implemented via external reactions (not belonging to the CRN) or via abundant species whose concentrations are negligibly affected by the CRN reactions within relevant time scales.
	An indirect control may be achieved via semipermeable membranes or by controlled injection of chemicals in continuous stirred-tank reactors.

Among the chemical species, the chemostatted ones are denoted by the indices $\sigma_{\mathrm{y}} \in \mathcal{S}_{\mathrm{y}}$, and the \emph{internal} ones by $\sigma_{\mathrm{x}} \in \mathcal{S}_{\mathrm{x}}$ ($\mathcal{S} \equiv \mathcal{S}_{\mathrm{x}} \cup \mathcal{S}_{\mathrm{y}}$).
Also, the part of the stoichiometric matrix related to the internal (resp. chemostatted) species is denoted by $\nabla^{\mathrm{X}} = \sset{ \nabla^{\sigma_{\mathrm{x}}}_{\rho} }$ (resp. $\nabla^{\mathrm{Y}} = \sset{ \nabla^{\sigma_{\mathrm{y}}}_{\rho} }$).

\begin{example}
	When chemostatting the CRN in fig.~\ref{fig:cCRN} as in fig.~\ref{fig:oCRN} the stoichiometric matrix \eqref{ex:smExample} splits into
	\begin{equation}
		\nabla^{\mathrm{X}} = 
		\begin{pmatrix}
			2 & 0 \\
			1 & -1 \\
			0 & -1
		\end{pmatrix}
		\, , \quad
		\nabla^{\mathrm{Y}} = 
		\begin{pmatrix}
			-1 & 0 \\
			0 & 1
		\end{pmatrix}
		\, .
		\label{}
	\end{equation}
	\qed
\end{example}

In \emph{nondriven open CRNs} the chemostatted species have constant concentrations, \emph{i.e.} $\left\{ \dt{Z}^{\sigma_{\mathrm{y}}} = 0 \right\}$.
In \emph{driven open CRNs} the chemostatted concentrations change over time according to some time-dependent protocol $\pi(t)$: $\left\{ Z^{\sigma_{\mathrm{y}}} \equiv Z^{\sigma_{\mathrm{y}}} (\pi(t)) \right\}$.
The changes of the internal species are solely due to reactions and satisfy the rate equations 
\begin{equation}
	\dt{Z}^{\sigma_{\mathrm{x}}} = \nabla^{\sigma_{\mathrm{x}}}_{\rho} J^{\rho} \, , \quad \forall \, \sigma_{\mathrm{x}} \in \mathcal{S}_{\mathrm{x}} \, .
	\label{ex:intflux}
\end{equation}
Instead, the changes of chemostatted species $\set{\de_{t} Z^{\sigma_{\mathrm{y}}}}$ are not only given by the \emph{species formation rates} $\set{\nabla^{\sigma_{\mathrm{y}}}_{\rho} J^{\rho}}$ but must in addition contain the \emph{external currents} $\left\{ I^{\sigma_{\mathrm{y}}} \right\}$, which quantify the rate at which chemostatted species enter into the CRN (negative if chemostatted species leave the CRN),
\begin{equation}
	\dt{Z}^{\sigma_{\mathrm{y}}} = \nabla^{\sigma_{\mathrm{y}}}_{\rho} J^{\rho} + I^{\sigma_\mathrm{y}} \, , \quad \forall \, \sigma_{\mathrm{y}} \in \mathcal{S}_{\mathrm{y}} \, .
	\label{ex:extflux}
\end{equation}
This latter equation is not a differential equation since the chemostatted concentrations $\set{Z^{\sigma_{\mathrm{y}}}}$ are not dynamical variables.
It shows that the external control of the chemostatted concentration is not necessarily direct, via the chemostatted concentrations, but can also be indirectly controlled via the external currents.
We note that \eqref{ex:extflux} is the dynamical expression of the decomposition of changes of species populations in internal--external introduced by de~Donder (see \cite[\S\S~4.1 and 15.2]{kondepudi14}).

A \emph{steady-state distribution} $\set{\bar{Z}^{\sigma_{\mathrm{x}}}}$, if it exists, must satisfy
\begin{subequations}
	\begin{align}
		\nabla^{\sigma_{\mathrm{x}}}_{\rho} \, \bar{J}^{\rho}  & = 0 \, , & & \forall \, \sigma_{\mathrm{x}} \in \mathcal{S}_{\mathrm{x}} \, , \label{eq:sfrSSi} \\
		\nabla^{\sigma_{\mathrm{y}}}_{\rho} \, \bar{J}^{\rho} + \bar{I}^{\sigma_{\mathrm{y}}} \label{eq:sfrSSf} & = 0  \, , & & \forall \, \sigma_{\mathrm{y}} \in \mathcal{S}_{\mathrm{y}} \, ,
	\end{align}
	\label{eq:sfrSS}
\end{subequations}
for given chemostatted concentrations $\set{Z^{\sigma_{\mathrm{y}}}}$.

\subsection{Conservation Laws}
\label{sec:cl}

In a \emph{closed} CRN, a \emph{conservation law} $\bm{\ell} = \set{\ell_{\sigma}}$ is a left null eigenvector of the stoichiometric matrix $\nabla$ \cite{alberty03,palsson06}
\begin{equation}
	\ell_{\sigma} \, \nabla^{\sigma}_{\rho} = 0 \, , \quad \forall \, \rho \in \mathcal{R} \, .
	\label{def:cl}
\end{equation}
Conservation laws identify conserved quantities $L \equiv \ell_{\sigma} \, Z^{\sigma}$, called \emph{components} \cite{alberty03,palsson06}, which satisfy
\begin{equation}
	\dt{L} = \ell_{\sigma} \, \dt{Z}^{\sigma} = 0 \, .
	\label{eq:component}
\end{equation}

We denote a set of independent conservation laws of the closed network by $\left\{ \bm{\ell}^{\lambda} \right\}$ and the corresponding components by $\left\{ L^{\lambda} \equiv \ell^{\lambda}_{\sigma} \, Z^{\sigma} \right\}$.
The choice of this set is not unique, and different choices have different physical meanings.
This set is never empty since the total mass is always conserved.
Physically, conservation laws are often related to parts of molecules, called \emph{moieties} \cite{goldbook:moiety}, which are exchanged between different species and/or subject to isomerization (see example \ref{exm:cl}).

In an \emph{open} CRN, since only $\set{Z^{\sigma_{\mathrm{x}}}}$ are dynamical variables, the conservation laws become the left null eigenvectors of the stoichiometric matrix of the internal species $\nabla^{\mathrm{X}}$.
Stated differently, when starting from the closed CRN, the chemostatting procedure may break a subset of the conservation laws of the closed network $\set{\bm\ell^{\lambda}}$ \cite{polettini14}.
\emph{E.g.} when the first chemostat is introduced the total mass conservation law is always broken.
Within the set $\left\{ \bm\ell^{\lambda} \right\}$, we label the broken ones by $\lambda_{\mathrm{b}}$ and the unbroken ones by $\lambda_{\mathrm{u}}$.
The broken conservation laws are characterized by
\begin{equation}
	\underbrace{\ell^{\lambda_{\mathrm{b}}}_{\sigma_{\mathrm{x}}} \, \nabla^{\sigma_{\mathrm{x}}}_{\rho}}_{\textstyle \neq 0} + \ell^{\lambda_{\mathrm{b}}}_{\sigma_{\mathrm{y}}} \, \nabla^{\sigma_{\mathrm{y}}}_{\rho} = 0 \, , \quad \forall \, \rho \in \mathcal{R} \, ,
	\label{}
\end{equation}
where the first term is nonvanishing for at least one $\rho \in \mathcal{R}$.
The \emph{broken components} $\set{L^{\lambda_{\mathrm{b}}} \equiv \ell^{\lambda_{\mathrm{b}}}_{\sigma} Z^{\sigma}}$ are no longer constant over time.
On the other hand, the unbroken conservation laws are characterized by
\begin{equation}
	\underbrace{\ell^{\lambda_{\mathrm{u}}}_{\sigma_{\mathrm{x}}} \, \nabla^{\sigma_{\mathrm{x}}}_{\rho}}_{\textstyle = 0} + \ell^{\lambda_{\mathrm{u}}}_{\sigma_{\mathrm{y}}} \, \nabla^{\sigma_{\mathrm{y}}}_{\rho} = 0 \, , \quad \forall \, \rho \in \mathcal{R} \, ,
	\label{}
\end{equation}
where the first term vanishes for all $\rho \in \mathcal{R}$.
Therefore, the \emph{unbroken components} $\set{L^{\lambda_{\mathrm{u}}} \equiv \ell^{\lambda_{\mathrm{u}}}_{\sigma} Z^{\sigma}}$ remain constant over time.
Without loss of generality, we choose the set $\{ \bm\ell^{\lambda} \}$ such that the entries related to the chemostatted species vanish, $\ell^{\lambda_{\mathrm{u}}}_{\sigma_{\mathrm{y}}} = 0, \, \forall \lambda_{\mathrm{u}}, \sigma_{\mathrm{y}}$.

\begin{figure}[t]
	\centering
	\includegraphics[width=.48\textwidth]{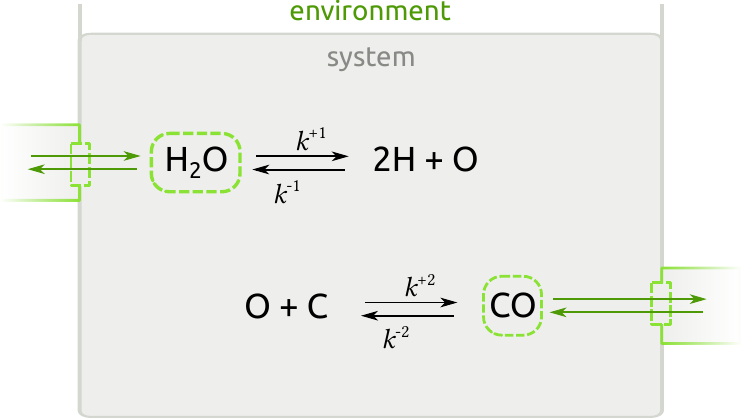}
	\caption{Specific implementation of the CRN in fig.~\ref{fig:oCRN}.}
	\label{fig:eCRN}
\end{figure}

\begin{example}
	\label{exm:cl}
	For the CRN in fig.~\ref{fig:cCRN}, an independent set of conservation laws is
	\begin{equation}
		\begin{split}
			\bm\ell^{1} & = 
			\begin{pmatrix}
				2 & 1 & 0 & 0 & 0
			\end{pmatrix} \, , \\
			\bm\ell^{2} & = 
			\begin{pmatrix}
				0 & 0 & 0 & 1 & 1
			\end{pmatrix} \, \text{and} \\
			\bm\ell^{3} & =
			\begin{pmatrix}
				0 & \tfrac{1}{2} & -1 & 1 & 0
			\end{pmatrix} \, .
		\end{split}
		\label{example:cl}
	\end{equation}
	When chemostatting as in fig.~\ref{fig:oCRN}, the first two conservation laws break while the last one remains unbroken.
	We also note that this set is chosen so that the unbroken conservation law satisfies $\ell^{3}_{\mathrm{a}} = \ell^{3}_{\mathrm{e}} = 0$.
	When considering the specific implementation in fig.~\ref{fig:eCRN} of the CRN in fig.~\ref{fig:oCRN}, we see that the first two conservation laws in \eqref{example:cl} represent the conservation of the concentrations of the moiety $\mathrm{H}$ and $\mathrm{C}$, respectively.
	Instead, the third conservation law in \eqref{example:cl} does not have a straightforward interpretation.
	It is related to the fact that when the species $\mathrm{H}$ or $\mathrm{C}$ are produced also $\mathrm{O}$ must be produced and \emph{vice versa}.
	\qed
\end{example}

\subsection{Detailed-Balanced Networks}
\label{sec:dbn}

A steady state \eqref{eq:sfrSS} is said to be an equilibrium state, $\{ Z_{\mathrm{eq}}^{\sigma} \}$, if it satisfies the \emph{detailed balance property} \cite[\S~9.4]{kondepudi14}, \emph{i.e.} all concentration currents \eqref{ex:currents} vanish,
\begin{equation}
	J^{\rho}_{\mathrm{eq}} \equiv J^{\rho} \big( \{ Z^{\sigma}_{\mathrm{eq}} \} \big) = 0, \quad \forall \, \rho \in \mathcal{R} \, .
	\label{def:db}
\end{equation}
For open networks, this means that the external currents, eq.~\eqref{eq:sfrSSf}, must also vanish $\set{I_{\mathrm{eq}}^{\sigma_{\mathrm{y}}} = 0}$.
By virtue of mass action kinetics, eq.~\eqref{ex:mak}, the detailed balance property \eqref{def:db} can be rewritten as
\begin{equation}
	\frac{k^{+\rho}}{k^{-\rho}} = {Z_{\mathrm{eq}}^{\sigma}}^{\nabla^{\rho}_{\sigma}} \, , \quad \forall \, \rho \in \mathcal{R} \, .
	\label{eq:db}
\end{equation}

A CRN is said to be \emph{detailed-balanced} if, for given kinetics $\set{k^{\pm \rho}}$ and chemostatting $\set{Z^{\sigma_{\mathrm{y}}}}$, its dynamics exhibits an equilibrium steady state \eqref{def:db}.
For each set of unbroken components $\set{L^{\lambda_{\mathrm{u}}}}$---which are given by the initial condition and constrain the space where the dynamics dwells---the equilibrium distribution is globally stable \cite{schuster89}.
Equivalently, detailed-balanced networks always relax to an equilibrium state, which for a given kinetics and chemostatting is unique and depends on the unbroken components only, see also sec.~\ref{sec:dbnThermo}.

Closed CRNs must be detailed-balanced.
This statement can be seen as the zeroth law for CRNs.
Consequently, rather than considering eq.~\eqref{eq:db} as a property of the equilibrium distribution, we impose it as a property that the rate constants must satisfy and call it \emph{local detailed-balance property}.
It is a universal property of elementary reactions which holds regardless of the network state.
Indeed, while the equilibrium distribution depends on the components, the r.h.s. of eq.~\eqref{eq:db} does not. This point will become explicit after introducing the thermodynamic structure, eq.~\eqref{ex:equilibrium} in sec.\ref{sec:dbnThermo}.
The local detailed-balance property will be rewritten in a thermodynamic form in \S~\ref{sec:affinities}, eq.~\eqref{ex:ldb}.

In open nondriven CRNs, the chemostatting procedure may prevent the system from reaching an equilibrium state.
To express this scenario algebraically we now introduce the concepts of emergent cycle and cycle affinity.

A cycle $\mathbf{\tilde{c}} = \set{\tilde{c}^{\rho}}$ is a right null eigenvector of the stoichiometric matrix \cite{polettini14}, namely
\begin{equation}
	\nabla^{\sigma}_{\rho} \, \tilde{c}^{\rho} = 0 \, , \quad \forall \, \sigma \in \mathcal{S} \, .
	\label{def:cycle}
\end{equation}
Since $\nabla$ is integer-valued, $\mathbf{\tilde{c}}$ can always be rescaled to only contain integer coefficients.
In this representation, its entries denote the number of times each reaction occurs (negative signs identify reactions occurring in backward direction) along a transformation which overall leaves the concentration distributions $\set{Z^{\sigma}}$ unchanged, see example \ref{exm:emergent}.
We denote by $\set{\mathbf{\tilde{c}}_{\alpha}}$ a set of linearly independent cycles.
An \emph{emergent cycle} $\mathbf{c} = \set{c^{\rho}}$ is defined algebraically as \cite{polettini14}
\begin{equation}
	\left\lbrace
	\begin{aligned}
		\nabla^{\sigma_{\mathrm{x}}}_{\rho} c^{\rho} & = 0 \, , \quad \forall \, \sigma_{\mathrm{x}} \in \mathcal{S}_{\mathrm{x}} \\
		\nabla^{\sigma_{\mathrm{y}}}_{\rho} c^{\rho} & \neq 0 \, , \quad \text{for at least one } \sigma_{\mathrm{y}} \in \mathcal{S}_{\mathrm{y}} \, .
	\end{aligned}
	\right.
	\label{def:emergent}
\end{equation}
In its integer-valued representation, the entries of $\mathbf{c}$ denote the number of times each reaction occurs along a transformation which overall leaves the concentrations of the internal species $\set{Z^{\sigma_{\mathrm{x}}}}$ unchanged while changing the concentrations of the chemostatted species by an amount $\nabla^{\sigma_{\mathrm{y}}}_{\rho} c^{\rho}$.
These latter are however immediately restored to their prior values due to the injection of $- \nabla^{\sigma_{\mathrm{y}}}_{\rho} c^{\rho}$ molecules of $\ce{X_{\sigma_{\mathrm{y}}}}$ performed by the chemostats.
Emergent cycles are thus pathways transferring chemicals across chemostats while leaving the internal state of the CRN unchanged.
We denote by $\set{\mathbf{c}_{\varepsilon}}$ a set of linearly independent emergent cycles.

When chemostatting an initially closed CRN, for each species which is chemostatted, either a conservation law breaks---as mentioned in \S~\ref{sec:cl}---or an independent emergent cycle arises \cite{polettini14}.
This follows from the rank nullity theorem for the stoichiometric matrices $\nabla$ and $\nabla^{\mathrm{X}}$, which ensures that the number of chemostatted species $|\mathcal{S}_{\mathrm{y}}|$ equals the number of broken conservation laws $|\lambda_{\mathrm{b}}|$ plus the number of independent emergent cycles $|\varepsilon|$: $|\mathcal{S}_{\mathrm{y}}| = |\lambda_{\mathrm{b}}| + |\varepsilon|$.
Importantly, the rise of emergent cycles is a topological feature:
it depends on the species which are chemostatted, but not on the chemostatted concentrations.
We also note that emergent cycles are modeled as ``flux modes'' in the context of metabolic networks \cite{leiser87,schuster94,klamt03}.

\begin{example}
	\label{exm:emergent}
	To illustrate the concepts of cycles and emergent cycles, we use the following CRN \cite{polettini14}
	\begin{equation}
		\includegraphics[width=.25\textwidth]{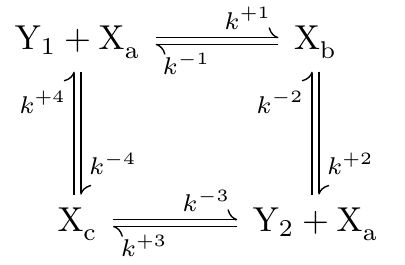}
		\label{tikz:matteo}
	\end{equation}
	whose $\ce{Y}_{1}$ and $\ce{Y}_{2}$ species are chemostatted.
	The stoichiometric matrix decomposes as
	\begin{equation}
		\hspace{-0.5cm}
		\begin{aligned}
			\nabla^{\mathrm{X}} = 
			& \begin{pmatrix}
				-1 & 1 & -1 & 1 \\
				1 & -1 & 0 & 0 \\
				0 & 0 & 1 & -1
			\end{pmatrix}
			&
			\nabla^{\mathrm{Y}} = 
			& \begin{pmatrix}
				-1 & 0 & 0 & 1 \\
				0 & 1 & -1 & 0
			\end{pmatrix}
			\, .
		\end{aligned}
	\end{equation}
	The set of linearly independent cycles \eqref{def:cycle} consists of only one cycle, which can be written as 
	\begin{equation}
		\tilde{\mathbf{c}} = 
		\begin{pmatrix}
			1 & 1 & 1 & 1 \\ 
		\end{pmatrix}\transpose \, .
		\label{eq:exampleCycle}
	\end{equation}
	As the CRN is chemostatted one linearly independent emergent cycle \eqref{def:emergent} arises
	\begin{equation}
		\mathbf{c} = 
		\begin{pmatrix}
			1 & 1 & - 1 & - 1 \\ 
		\end{pmatrix}\transpose \, .
		\label{eq:exampleEmergentCycle}
	\end{equation}
	We now see that if each reaction occurs a number of times given by the entry of the cycle \eqref{eq:exampleCycle}, the CRN goes back to the initial state, no matter which one it is.
	On the other hand, when the emergent cycle \eqref{eq:exampleEmergentCycle} is performed, the state of the internal species does not change, while two molecules of $\ce{Y_{1}}$ are annihilated and two of $\ce{Y_{2}}$ are created.
	However, since the chemostats restore their initial values, the overall result of $\mathbf{c}$ is to transfer two $\ce{Y_{1}}$, transformed in $\ce{Y_{2}}$, from the first to the second chemostat.

	The closed version of this CRN has two independent conservation laws,
	\begin{equation}
		\begin{aligned}
			\bm \ell^{1} & = 
			\begin{pmatrix}
				0 & 1 & 1 & 1 & 1 \\ 
			\end{pmatrix} \\
			\bm \ell^{2} & = 
			\begin{pmatrix}
				1 & 1 & 1 & 0 & 0 \\ 
			\end{pmatrix}
		\end{aligned}
		\label{}
	\end{equation}
	the first of which, $\bm \ell^{1}$, is broken following the chemostatting of any of the two species $\ce{Y_{1}}$ or $\ce{Y_{2}}$.
	The other chemostatted species, instead, gives rise to the emergent cycle \eqref{eq:exampleEmergentCycle}, so that the relationship $|\mathcal{S}_{\mathrm{y}}| = |\lambda_{\mathrm{b}}| + |\varepsilon|$ is satisfied.
	\qed
\end{example}

Any cycle $\mathbf{\tilde{c}}_{\alpha}$ and emergent cycle $\mathbf{c}_{\varepsilon}$ bears a \emph{cycle affinity} \cite{polettini14}
\begin{align}
	\mathcal{\tilde{A}}_{\alpha} & =  \tilde{c}_{\alpha}^{\rho} \, RT \ln \frac{J_{+\rho}}{J_{-\rho}} \, , \\
	\mathcal{A}_{\varepsilon} & =  c_{\varepsilon}^{\rho} \, RT \ln \frac{J_{+\rho}}{J_{-\rho}} \, .
	\label{ex:dynamical}
\end{align}
From the definition of cycle \eqref{def:cycle} and current \eqref{ex:currents}, and the local detailed balance \eqref{eq:db}, it follows that the cycle affinities along the cycles \eqref{def:cycle} vanish, $\set{\mathcal{\tilde{A}}_{\alpha} = 0}$, and that the cycle affinities along the emergent cycles only depend on the chemostatted concentrations
\begin{equation}
	\mathcal{A}_{\varepsilon} = c_{\varepsilon}^{\rho} \, RT \ln \frac{k_{+\rho}}{k_{-\rho}} Z_{\sigma_{\mathrm{y}}}^{- \nabla^{\sigma_{\mathrm{y}}}_{\rho}} \, .
	\label{ex:affinity}
\end{equation}
Since emergent cycles are pathways connecting different chemostats, the emergent affinities quantify the chemical forces acting along the cycles.
This point will become clearer later, when the thermodynamic expressions of the emergent cycle affinities $\set{\mathcal{A}_{\varepsilon}}$ will be given, eq.~\eqref{ex:affinityG}.

A CRN is detailed-balanced if and only if all the emergent cycle affinities $\set{\mathcal{A}_{\varepsilon}}$ vanish.
This condition is equivalent to the \emph{Wegscheider's condition} \cite{schuster89}.
This happens when the chemostatted concentrations fit an equilibrium distribution.
As a special case, \emph{unconditionally detailed-balanced networks} are open CRNs with no emergent cycle.
Therefore, they are detailed-balanced for any choice of the chemostatted concentrations.
Consequently, even when a time-dependent driving acts on such a CRN and prevents it from reaching an equilibrium state, a well-defined equilibrium state exists at any time:
the equilibrium state to which the CRN would relax if the time-dependent driving were stopped.

\begin{example}
	Any CRN with one chemostatted species only ($|\mathcal{S}_{\mathrm{y}}|=1$) is unconditionally detailed-balanced.
	Indeed, as mentioned in \S~\ref{sec:cl}, the first chemostatted species always breaks the mass conservation law, $|\lambda_{\mathrm{b}}|=1$ and thus no emergent cycle arises, $|\varepsilon| = |\mathcal{S}_{\mathrm{y}}| - |\lambda_{\mathrm{b}}| = 0$.

	The open CRN in fig.~\ref{fig:oCRN} is an example of unconditionally detailed-balanced network with two chemostatted species, since the chemostatting breaks two conservation laws, see example~\ref{exm:cl}.
	Indeed, a nonequilibrium steady state would require a continuous injection of $\ce{Y}_{\mathrm{a}}$ and ejection of $\ce{Y}_{\mathrm{e}}$ (or \emph{vice versa}).
	But this would necessary result in a continuous production of $\ce{X}_{\mathrm{b}}$ and consumption of $\ce{X}_{\mathrm{d}}$ which is in contradiction with the steady state assumption.
	\qed
\end{example}

Finally, a tacit assumption in the above discussion is that the network involves a finite number of species and reactions, \emph{i.e.} the CRN is finite-dimensional.
Infinite-dimensional CRNs can exhibits long-time behaviors different from equilibrium even in absence of emergent cycles \cite{rao15:denzymes}.

\subsection{Complex-Balanced Networks}
\label{sec:cbn}

To discuss complex-balanced networks and complex-balanced distributions, we first introduce the notion of complex in open CRNs.

A \emph{complex} is a group of species which combines in a reaction as products or as reactants. 
Each side of eq.~\eqref{reaction} defines a complex but different reactions might involve the same complex.
We label complexes by $\gamma \in \mathcal{C}$, where $\mathcal{C}$ is the set of complexes.

\begin{example}
	\label{exm:cbClosed}
	Let us consider the following CRN \cite{horn73:stability}
	\begin{equation}
		\begin{aligned}
			& \ce{X_{\mathrm{a}} <=>[k^{+1}][k^{-1}] X_{\mathrm{b}}} \\
			& \ce{X_{\mathrm{a}} + X_{\mathrm{b}} <=>[k^{+2}][k^{-2}] 2X_{\mathrm{b}} <=>[k^{+3}][k^{-3}] X_{\mathrm{c}}}
		\end{aligned} \, .
		\label{ce:complexbalanced}
	\end{equation}
	The set of complexes is $\mathcal{C}=\set{\ce{X}_{\mathrm{a}}, \ce{X}_{\mathrm{b}}, \ce{X}_{\mathrm{a}} + \ce{X}_{\mathrm{b}}, 2\ce{X}_{\mathrm{b}}, \ce{X}_{\mathrm{c}}}$, and the complex $2\ce{X}_{\mathrm{b}}$ is involved in both the second and third reaction.
	\qed
\end{example}

The notion of complex allows us to decompose the stoichiometric matrix $\nabla$ as
\begin{equation}
	\nabla^{\sigma}_{\rho} = \Gamma^{\sigma_{\mathrm{}}}_{\gamma} \, \partial^{\gamma}_{\rho} \, .
	\label{ex:NablaDec}
\end{equation}
We call $\Gamma = \set{\Gamma^{\sigma}_{\gamma}}$ the \emph{composition matrix} \cite{horn72,horn72:complex}.
Its entries $\Gamma^{\sigma}_{\gamma}$ are the stoichiometric number of species $\ce{X}_{\sigma}$ in the complex $\gamma$.
The composition matrix encodes the structure of each complex in terms of species, see example \ref{ex:compositionAndIncidence}.
The matrix $\partial = \set{\partial^{\gamma}_{\rho}}$ denotes the \emph{incidence matrix} of the CRN, whose entries are given by
\begin{equation}
	\partial^{\gamma}_{\rho} = 
	\begin{cases}
		1 & \text{if $\gamma$ is the product complex of $+\rho$} \, , \\
		-1 & \text{if $\gamma$ is the reactant complex of $+\rho$} \, , \\
		0 & \text{otherwise.}
	\end{cases}
	\label{def:incidence}
\end{equation}
The incidence matrix encodes the structure of the network at the level of complexes, \emph{i.e.} how complexes are connected by reactions.
If we think of complexes as network nodes, the incidence matrix associates an edge to each reaction pathway and the resulting topological structure is a \emph{reaction graph}, \emph{e.g.} fig.~\ref{fig:cCRN} and eqs.~\eqref{tikz:matteo} and \eqref{ce:complexbalanced}.
The stoichiometric matrix instead encodes the structure of the network at the level of species.
If we think of species as the network nodes, the stoichiometric matrix does not define a graph since reaction connects more than a pair of species, in general.
The structure originating is rather a \emph{hyper-graph} \cite{klamt09,polettini14}, or equivalently a \emph{Petri net} \cite{petri08,baez14}.

\begin{example}
	\label{ex:compositionAndIncidence}
	The composition matrix and the incidence matrix of the CRN in \eqref{ce:complexbalanced} are
	\begin{equation}
		\begin{aligned}
			\Gamma & = 
			\begin{pmatrix}
				1 & 0 & 1 & 0 & 0 \\
				0 & 1 & 1 & 2 & 0 \\
				0 & 0 & 0 & 0 & 1 
			\end{pmatrix} \, , &
			\partial & = 
			\begin{pmatrix}
				-1	& 0		&  0 \\
				1	& 0		&  0 \\
				0	& -1	&  0 \\
				0	& 1		& -1 \\
				0	& 0		&  1 
			\end{pmatrix} \, ,
		\end{aligned}
		\label{}
	\end{equation}
	where the complexes are ordered as in example~\ref{exm:cbClosed}.
	The corresponding reaction hyper-graph is
	\begin{equation}
		\includegraphics[width=.3\textwidth]{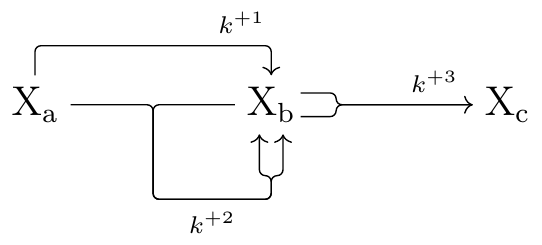}
	\end{equation}
	where only the forward reactions are depicted.
	\qed
\end{example}

In an open CRN, we regroup all complexes $\gamma \in \mathcal{C}$ of the closed CRN which have the same stoichiometry for the internal species (\emph{i.e.} all complexes with the same internal part of the composition matrix $\Gamma^{\mathrm{X}}_{\gamma}$ regardless of the chemostatted part $\Gamma^{\mathrm{Y}}_{\gamma}$) in sets denoted by $\mathcal{C}_{j}$, for $j=1,2,\dots$.
Complexes of the closed network made solely of chemostatted species in the open CRN are all regrouped in the same complex $\mathcal{C}_0$.
This allows to decompose the internal species stoichiometric matrix as
\begin{equation}
	\nabla^{\sigma_{\mathrm{x}}}_{\rho} = \Gamma^{\sigma_{\mathrm{x}}}_{j} \, \partial^{j}_{\rho} \, .
	\label{}
\end{equation}
where $\set{\Gamma^{\sigma_{\mathrm{x}}}_{j} \equiv \Gamma^{\sigma_{\mathrm{x}}}_{\gamma}, \text{ for } \gamma \in \mathcal{C}_{j}}$ are the entries of the composition matrix corresponding to the internal species, and $\set{\partial^{j}_{\rho} \equiv \sum_{\gamma \in \mathcal{C}_{j}} \partial^{\gamma}_{\rho}}$ are the entries of the incidence matrix describing the network of regrouped complexes.
This regrouping corresponds to the---equivalent---CRN only made of internal species with \emph{effective rate constant} $\set{k^{\pm \rho} {Z^{\sigma_{\mathrm{y}}}}^{\nabla^{\pm \rho}_{\sigma_{\mathrm{y}}}}}$ ruling each reaction.

\begin{example}
	Let us consider the CRN \eqref{ce:complexbalanced} where the species $\ce{X}_{\mathrm{a}}$ and $\ce{X}_{\mathrm{c}}$ are chemostatted.
	The five complexes of the closed network, see example \ref{exm:cbClosed}, are regrouped as $\mathcal{C}_0=\{\ce{X}_{\mathrm{a}}, \ce{X}_{\mathrm{c}}\}$, $\mathcal{C}_1=\{\ce{X}_{\mathrm{b}}, \ce{X}_{\mathrm{b}} + \ce{X}_{\mathrm{a}}\}$ and $\mathcal{C}_2=\{\ce{2 X_{\mathrm{b}}}\}$.
	In terms of these groups of complexes, the composition matrix and incidence matrix are
	\begin{equation}
		\begin{aligned}
			\Gamma^{\mathrm{X}} & = 
			\begin{pmatrix}
				0 & 1 & 2
			\end{pmatrix} \, , &
			\partial^{\mathcal{C}} & = 
			\begin{pmatrix}
				-1	& 0		&  1 \\
				1	& -1	&  0 \\
				0	& 1		& -1
			\end{pmatrix} \, ,
		\end{aligned}
		\label{}
	\end{equation}
	which corresponds to the effective representation
	\begin{equation}
		\label{cn:cbOpen}
		\includegraphics[width=.22\textwidth]{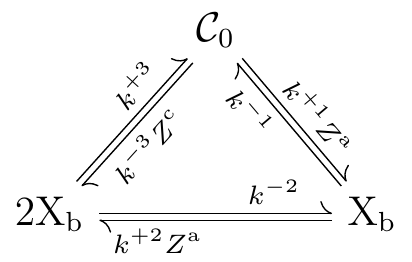}
	\end{equation}
	\qed
\end{example}

A steady-state distribution $\set{\bar{Z}^{\sigma_{\mathrm{x}}}}$ \eqref{eq:sfrSS} is said to be \emph{complex-balanced} if the net current flowing in each group of complexes $\mathcal{C}_{j}$ vanishes, \emph{i.e.} if the currents $\set{\bar{J}^{\rho}}$ satisfy
\begin{equation}
	\partial^{j}_{\rho} \bar{J}^{\rho} \equiv \sum_{\gamma \in \mathcal{C}_{j}} \partial^{\gamma}_{\rho} \, \bar{J}^{\rho} = 0 \, , \quad \forall j \, .
	\label{def:cb}
\end{equation}
Complex-balance steady states are therefore a subclass of steady states \eqref{eq:sfrSSi} which include equilibrium ones \eqref{def:db} as a special case
\begin{equation}
	\underbrace{\Gamma^{\sigma_{\mathrm{x}}}_{j} \quad \; \underbrace{\partial^{j}_{\rho} \underbrace{\bar{J}^{\rho}}
	_{\textstyle = 0 \text{ iff DB}}}
	_{\textstyle = 0 \text{ iff CB}}}
	_{\textstyle = 0 \text{ for generic SS}} \, .
	\label{}
\end{equation}
While for generic steady states only the \emph{internal species formation rates} vanish, for complex-balanced ones the \emph{complex formation rates} also vanish.

For a fixed kinetics ($\set{k^{\pm \rho}}$) and chemostatting ($\mathcal{S}_{\mathrm{y}}$ and $\set{Z^{\sigma_{\mathrm{y}}}}$), a CRN is \emph{complex-balanced} if its dynamics exhibits a complex-balanced steady state \eqref{def:cb} \cite{horn72,feinberg72}.
The complex-balanced distribution \eqref{def:cb} depends on the unbroken components $\set{L^{\lambda_{\mathrm{u}}}}$, which can be inferred from the initial conditions, and is always globally stable \cite{craciun15}.
Hence, complex-balanced networks always relax to a---complex-balanced---steady state.
Detailed-balanced networks are a subclass of complex-balanced networks.

Whether or not a CRN is complex-balanced depends on the network topology ($\nabla$), the kinetics ($\set{k^{\pm \rho}}$) and the chemostatting ($\mathcal{S}_{\mathrm{y}}$ and $\set{Z^{\sigma_{\mathrm{y}}}}$).
For any given network topology and set of chemostatted species $\mathcal{S}_{\mathrm{y}}$, one can always find a set of effective rate constants $\set{k^{\pm \rho} {Z^{\sigma_{\mathrm{y}}}}^{\nabla^{\pm \rho}_{\sigma_{\mathrm{y}}}}}$ which makes that CRN complex-balanced \cite{horn72:complex}.
However, for some CRNs, this set coincides with the one which makes the CRN detailed-balanced \cite{dickenstein11}.
A characterization of the set of effective rate constants which make a CRN complex balanced is reported in Refs.~\cite{horn72:complex,dickenstein11}.

\emph{Deficiency-zero} CRNs are a class of CRN which is complex-balanced irrespective of the effective kinetics $\set{k^{\pm \rho} {Z^{\sigma_{\mathrm{y}}}}^{\nabla^{\pm \rho}_{\sigma_{\mathrm{y}}}}}$ \cite{horn72,feinberg72,horn72:complex}.
The \emph{network deficiency} is a topological property of the CRN which we briefly discuss in app.~\ref{sec:dzn}, see Refs.~\cite{feinberg:lectures,gunawardena03,polettini15} for more details.
Consequently, regardless of the way in which a deficiency-zero CRN is driven in time, it will always remain complex-balanced.
Throughout the paper, we will refer to these CRNs as \emph{unconditionally complex-balanced}, as in the seminal work \cite{horn72}.

\begin{example}
	\label{exm:cbSS}
	The open CRN \eqref{cn:cbOpen} has a single steady state $\bar{Z}^{\mathrm{b}}$ for any given set of rate constants and chemostatted concentrations $Z^{\mathrm{a}}$ and $Z^{\mathrm{c}}$ \cite{horn73:stability}, defined by \eqref{eq:sfrSSi}
	\begin{equation}
		\begin{split}
			\dt \bar{Z}^{\mathrm{b}} & =
			\bar{J}^{1} + \bar{J}^{2} - 2 \bar{J}^{3} \\
			& = k^{+1} Z^{\mathrm{a}} - k^{-1} \bar{Z}^{\mathrm{b}} + k^{+2} Z^{\mathrm{a}} \bar{Z}^{\mathrm{b}} - k^{-2} (\bar{Z}^{\mathrm{b}})^2 + \\
			& \quad + 2 k^{-3} Z^{\mathrm{c}} - 2 k^{+3} (\bar{Z}^{\mathrm{b}})^2  = 0 \, .
		\end{split}
		\label{}
	\end{equation}
	If the stronger condition \eqref{def:cb} holds
	\begin{equation}
		\begin{aligned}
			\bar{J}^{3} - \bar{J}^{1} & = 0 \, , \quad \text{(group } \mathcal{C}_0) \, , \\
			\bar{J}^{1} - \bar{J}^{2} & = 0 \, , \quad \text{(group } \mathcal{C}_1) \, , \\
			\bar{J}^{2} - \bar{J}^{3} & = 0 \, , \quad \text{(group } \mathcal{C}_2) \, ,
		\end{aligned}
		\label{eq:cbExample}
	\end{equation}
	which is equivalent to
	\begin{multline}
			k^{+1} Z^{\mathrm{a}} - k^{-1} \bar{Z}^{\mathrm{b}} =
			k^{+2} Z^{\mathrm{a}} \bar{Z}^{\mathrm{b}} - k^{-2} (\bar{Z}^{\mathrm{b}})^2 \\
			= k^{+3} (\bar{Z}^{\mathrm{b}})^2 - k^{-3} Z^{\mathrm{c}} \, ,
		\label{eq:cbExample2}
	\end{multline}
	the steady state is complex-balanced.
	Yet, if the steady-state currents are all independently vanishing,
	\begin{equation}
		\begin{aligned}
			\bar{J}^{1} = \bar{J}^{2} = \bar{J}^{3} = 0 \, ,
		\end{aligned}
		\label{}
	\end{equation}
	\emph{i.e.}, eq.~\eqref{eq:cbExample2} is equal to zero, then the steady state is detailed-balanced.

	When, for simplicity, all rate constants are taken as 1, the complex-balanced set of quadratic equations \eqref{eq:cbExample2} admits a positive solution $\bar{Z}^{\mathrm{b}}$ \emph{only if} $Z^{\mathrm{a}} = 2 - Z^{\mathrm{c}}$ ($0 < Z^{\mathrm{c}} < 2$) or $Z^{\mathrm{a}} = \sqrt{Z^{\mathrm{c}}}$.
	The former case corresponds to a genuine complex-balanced state, $\bar{Z}^{\mathrm{b}} = 1$ with currents $\bar{J}^{1} = \bar{J}^{2} = \bar{J}^{3} = 1 - Z^{\mathrm{c}}$, while the second to a detailed-balance state $\bar{Z}^{\mathrm{b}} = \sqrt{Z^{\mathrm{c}}}$ with vanishing currents.
	When, for example, $Z^{\mathrm{a}} = 1$ and $Z^{\mathrm{c}} = 4$, neither of the two previous conditions holds: the nonequilibrium steady state is $\bar{Z}^{\mathrm{b}} = \sqrt{3}$ with currents $\bar{J}^{1} = 1 - \sqrt{3}$, $\bar{J}^{2} = -3 + \sqrt{3}$, and $\bar{J}^{3} = -1$.
	\qed
\end{example}

\begin{example}
	Let us now consider the following open CRN \cite{polettini15}
	\begin{equation}
		\ce{Y_{\mathrm{a}} <=>[k^{+1}][k^{-1}] X_{\mathrm{b}} <=>[k^{+2}][k^{-2}] X_{\mathrm{c}} + X_{\mathrm{d}} <=>[k^{+3}][k^{-3}] Y_{\mathrm{e}}}
		\label{ce:enzyme}
	\end{equation}
	where the species $\ce{Y}_{\mathrm{a}}$ and $\ce{Y}_{\mathrm{e}}$ are chemostatted.
	Out of the four complexes of the closed network, $\{\ce{Y}_{\mathrm{a}},\ce{X}_{\mathrm{b}},\ce{X}_{\mathrm{c}}+\ce{X}_{\mathrm{d}},\ce{Y}_{\mathrm{e}}\}$, two are grouped into $\mathcal{C}_0=\{\ce{Y}_{\mathrm{a}},\ce{Y}_{\mathrm{e}}\}$ and the other two remain $\mathcal{C}_1=\set{\ce{X}_{\mathrm{b}}}$ and $\mathcal{C}_2=\set{\ce{X}_{\mathrm{c}}+\ce{X}_{\mathrm{d}}}$.
	The effective representation of this open CRN is
	\begin{equation}
		\includegraphics[width=.3\textwidth]{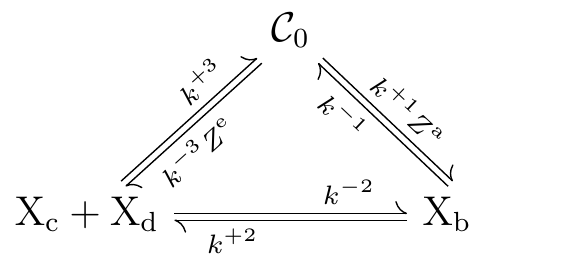}
	\end{equation}
	This network is deficiency-zero and hence unconditionally complex-balanced \cite{polettini15}.
	Therefore, given any set of rate constants $k^{\pm 1}$, $k^{\pm 2}$, and $k^{\pm 3}$, and the chemostatted concentrations $Z^{\mathrm{a}}$ and $Z^{\mathrm{e}}$, the steady state of this CRN is complex-balanced, \emph{i.e.} the steady state always satisfies a set of condition like those in eq.~\eqref{eq:cbExample}.
	Indeed, contrary to example \ref{exm:cbSS}, steady state currents $\set{\bar{J}^{1},\bar{J}^{2},\bar{J}^{3}}$ different from each other cannot exist since they would induce a growth or decrease of some concentrations.
	\qed
\end{example}

\section{Thermodynamics of\\Chemical Networks}
\label{sec:thermodynamics}

Using local equilibrium, we here build the connection between the dynamics and the nonequilibrium thermodynamics for arbitrary CRNs.
In the spirit of Stochastic Thermodynamics, we derive an energy and entropy balance, and express the dissipation of the CRN as the difference between the chemical work done by the reservoirs on the CRN and its change in nonequilibrium free energy. We finally discuss the information-theoretical content of the nonequilibrium free energy and its relation to the dynamical potentials used in Chemical Reaction Network Theory. 

\subsection{Local Equilibrium}
\label{sec:le}

Since we consider homogeneous reaction mixtures in \emph{ideal dilute solutions}, the assumption of \emph{local equilibrium} \cite[\S~15.1]{kondepudi14} \cite{prigogine49} means that the equilibration following any reaction event is much faster than any reaction time scale.
Thus, what is assumed is that the nonequilibrium nature of the thermodynamic description is solely due to the reaction mechanisms.
If all reactions could be instantaneously shut down, the state of the whole CRN would immediately become an equilibrated ideal mixture of species.
As a result, all the intensive thermodynamic variables are well-defined and equal everywhere in the system.
The temperature $T$ is set by the solvent, which acts as a thermal bath, while the pressure $p$ is set by the environment the solution is exposed to.
As a result, each chemical species is characterized by a \emph{chemical potential} \cite[\S~3.1]{alberty03}
\begin{equation}
	\mu_{\sigma} = \mu_{\sigma}^{\circ} + RT \ln \frac{Z_{\sigma}}{Z_{\mathrm{tot}}} \, , \quad \forall \, \sigma \in \mathcal{S} \, ,
	\label{ex:cp}
\end{equation}
where $R$ denotes the gas constant and $\set{\mu_{\sigma}^{\circ} \equiv \mu_{\sigma}^{\circ}(T)}$ are the standard-state chemical potentials, which depend on the temperature and on the nature of the solvent.
The total concentration of the solution is denoted by $Z_{\mathrm{tot}} = \sum_{\sigma} Z^{\sigma} + Z^{0}$, where $Z^{0}$ is the concentration of the solvent.
We assume for simplicity that the solvent does not react with the solutes.
In case it does, our results still hold provided one treats the solvent as a nondriven chemostatted species, as discussed in app.~\ref{app:idl}.
Since the solvent is much more abundant than the solutes, the total concentration is almost equal to that of the solvent which is a constant, $Z_{\mathrm{tot}} \simeq Z_{0}$.
Without loss of generality, the constant term $- RT \ln Z_{\mathrm{tot}} \simeq - RT \ln Z^{0}$ in eq.~\eqref{ex:cp} is absorbed in the standard-state chemical potentials.
Consequently, many equations appear with non-matching dimensions.
We also emphasize that standard state quantities, denoted with ``$^\circ$'', are defined as those measured in ideal conditions, at standard pressure ($p^\circ = 100$ kPa) and molar concentration ($Z^\circ_{\sigma} = 1$ mol/dm$^{3}$), but not at a standard temperature \cite[p.~61]{greenbook}.

Due to the assumption of local equilibrium and homogeneous reaction mixture, the densities of all extensive thermodynamic quantities are well-defined and equal everywhere in the system.
With a slight abuse of notation, we use the same symbol and name for densities as for their corresponding extensive quantity.
\emph{E.g.} $S$ is the molar entropy divided by the volume of the solution, but we denote it as \emph{entropy}.
We apply the same logic to rates of change.
\emph{E.g.} we call entropy production rate the molar entropy production density rate.

\subsection{Affinities, Emergent Affinities and\\Local Detailed Balance}
\label{sec:affinities}

The thermodynamic forces driving reactions are given by differences of chemical potential \eqref{ex:cp}
\begin{equation}
	\dr{G} \equiv \nabla^{\sigma}_{\rho} \, \mu_{\sigma} \, ,
\label{ex:gfer}
\end{equation}
also called \emph{Gibbs free energies of reaction} \cite[\S~9.3]{kondepudi14} \cite[\S~3.2]{alberty03}.
Since these must all vanish at equilibrium, $\nabla^{\sigma}_{\rho} \, \mu^{\mathrm{eq}}_{\sigma} = 0, \, \forall \rho$, we have
\begin{equation}
	\dr{G} = - RT \, \nabla^{\sigma}_{\rho} \, \ln \frac{Z_{\sigma}}{Z^{\mathrm{eq}}_{\sigma}} \, .
\end{equation}
The local detailed-balance \eqref{eq:db} allows us to express these thermodynamic forces in terms of \emph{reaction affinities},
\begin{equation}
	A_{\rho}
	\equiv RT \ln \frac{J_{+\rho}}{J_{-\rho}}
	= - \dr{G}
\label{ex:gferJ} 
\end{equation}
which quantify the kinetic force acting along each reaction pathway \cite[\S~4.1.3]{kondepudi14}.

The change of Gibbs free energy along emergent cycles,
\begin{equation}
	\mathcal{A}_{\varepsilon} = - c_{\varepsilon}^{\rho} \, \dr{G} = - c_{\varepsilon}^{\rho} \, \nabla^{\sigma_{\mathrm{y}}}_{\rho} \mu_{\sigma_{\mathrm{y}}} \, ,
	\label{ex:affinityG}
\end{equation}
gives the external thermodynamic forces the network is coupled to, as we shall see in eq. (\ref{ex:eprSS}), and thus provide a thermodynamic meaning to the cycle affinities \eqref{ex:affinity}.

Combining the detailed-balance property \eqref{eq:db} and the equilibrium condition on the affinities $A^{\mathrm{eq}}_{\rho} = 0$ \eqref{ex:gfer}, we can relate the Gibbs free energies of reaction to the rate constants
\begin{equation}
	\frac{k^{+\rho}}{k^{-\rho}} = \exp \left\{ -\frac{\drs{G}}{RT} \right\} \, ,
	\label{ex:ldb}
\end{equation}
where $\drs{G} \equiv \nabla^{\sigma}_{\rho} \, \mu^\circ_{\sigma}$.
This relation is the thermodynamic counterpart of the \emph{local detailed balance} \eqref{eq:db}.
It plays the same role as in Stochastic Thermodynamics, namely connecting the thermodynamic description to the stochastic dynamics.
We emphasize that the local detailed balance property as well as the local equilibrium assumption by no mean imply that the CRN operates close to equilibrium. 
Their importance is to assign well defined equilibrium potentials to the states of the CRN, which are then connected by the nonequilibrium mechanisms, \emph{i.e.} reactions. 

\subsection{Enthalpies and Entropies of Reaction}
\label{sec:EEreaction}

To identify the heat produced by the CRN, we need to distinguish the enthalpic change produced by each reaction from the entropic one.
We consider the decomposition of the standard state chemical potentials \cite[\S~3.2]{alberty03}:
\begin{equation}
	\mu^\circ_{\sigma} = h^\circ_{\sigma} - T s^\circ_{\sigma} \, .
	\label{ex:scpDEC}
\end{equation}
The \emph{standard enthalpies of formation} $\set{h^\circ_{\sigma}}$ take into account the enthalpic contributions carried by each species \cite[\S~3.2]{alberty03} \cite[\S~10.4.2]{holyst12}.
Enthalpy changes caused by reactions give the \emph{enthalpies of reaction} \cite[\S~3.2]{alberty03} \cite[\S~2.4]{kondepudi14}
\begin{equation}
	\dr{H} = \nabla^{\sigma}_{\rho} \, h^\circ_{\sigma} \, ,
	\label{ex:hor}
\end{equation}
which at constant pressure measure the \emph{heat of reaction}.
This is the content of the Hess' Law (see \emph{e.g.} \cite[\S~10.4.1]{holyst12}).
The \emph{standard entropies of formation} $\set{s^\circ_{\sigma}}$ take into account the internal entropic contribution carried by each species under standard-state conditions \cite[\S~3.2]{alberty03}.
Using \eqref{ex:scpDEC}, the chemical potentials (\ref{ex:cp}) can be rewritten as
\begin{equation}
	\mu_{\sigma} = h^\circ_{\sigma} - T \underbrace{\left( s^\circ_{\sigma} - R \ln Z_{\sigma} \right)}_{\displaystyle \equiv s_{\sigma}} \, .
	\label{ex:cpDEC}
\end{equation}
The \emph{entropies of formation} $\set{s_{\sigma} \equiv s^\circ_{\sigma} - R \ln Z_{\sigma}}$ account for the entropic contribution of each species in the CRN \cite[\S~3.2]{alberty03}.
Entropy changes along reactions are given by
\begin{equation}
	\dr{S} = \nabla^{\sigma}_{\rho} \, s_{\sigma} \, ,
	\label{ex:eor}
\end{equation}
called \emph{entropies of reaction} \cite[\S~3.2]{alberty03}.

\subsection{Entropy Balance}
\label{sec:entropyBalance}

\subsubsection{Entropy Production Rate}
\label{sec:epr}

The \emph{entropy production rate} is a non-negative measure of the break of detailed balance in each chemical reaction.
Its typical form is given by \cite[\S~9.5]{kondepudi14} \cite{schnakenberg76}
\begin{equation}
	T \epr \equiv RT \, (J_{+\rho}-J_{-\rho}) \, \ln \frac{J_{+\rho}}{J_{-\rho}} \geq 0 \, ,
	\label{ex:eprJ}
\end{equation}
because 
\begin{enumerate}
	\item It is non-negative and vanishes only at equilibrium, \emph{i.e.} when the detailed balance property \eqref{def:db} is satisfied;
	\item It vanishes to first order around equilibrium, thus allowing for quasi-static reversible transformations. Indeed, defining
	\begin{equation}
		\frac{{Z}^{\sigma} - Z_{\mathrm{eq}}^{\sigma}}{Z_{\mathrm{eq}}^{\sigma}} = {\epsilon}^{\sigma} \, , \quad \abs{{\epsilon}^{\sigma}} \ll 1 \, , \quad \forall \, \sigma \in \mathcal{S} \, ,
		\label{}
	\end{equation}
	we find that
	\begin{equation}
		\epr = E^{\sigma}_{\sigma'} \, \epsilon^{\sigma'} \, \epsilon_{\sigma} + \Order{\epsilon^{3}} \, ,
		\label{}
	\end{equation}
	where ${E} \equiv \set{E^{\sigma}_{\sigma'}}$ is a positive semidefinite symmetric matrix.
\end{enumerate}
Furthermore it can be rewritten in a thermodynamically appealing way using \eqref{ex:gferJ}:
\begin{equation}
	T \epr = - J^{\rho} \, \dr{G} \, .
\label{ex:eprG} 
\end{equation}
It can be further expressed as the sum of two distinct contributions \cite{polettini14}:
\begin{equation}
	T \epr = \underbrace{- \mu_{\sigma_{\mathrm{x}}} \dt{Z}^{\sigma_{\mathrm{x}}} }_{\textstyle \equiv T \vepr{\mathrm{x}}} \underbrace{- \mu_{\sigma_{\mathrm{y}}} \left( \dt{Z}^{\sigma_{\mathrm{y}}} - I^{\sigma_{\mathrm{y}}} \right) }_{\textstyle \equiv T \vepr{\mathrm{y}}} \, .
	\label{ex:erpIED}
\end{equation}
The first term is due to changes in the internal species and thus vanishes at steady state.
The second term is due to the chemostats.
It takes into account both the exchange of chemostatted species and the time-dependent driving of their concentration.
If the system reaches a nonequilibrium steady state, the external currents $\set{\bar{I}^{\sigma_{\mathrm{y}}}}$ do not vanish and the entropy production reads
\begin{equation}
	T \bar{\dot{S}}_{\mathrm{i}} = \bar{I}^{\sigma_{\mathrm{y}}} \, {\mu}_{\sigma_{\mathrm{y}}} \, .
	\label{}
\end{equation}
This expression can be rewritten as a bilinear form of emergent cycle affinities $\set{\mathcal{A}_{\varepsilon}}$ \eqref{ex:affinityG} and currents along emergent cycle $\set{\bar{\mathcal{J}}^{\varepsilon} \equiv c^{\varepsilon}_{\rho} \bar{J}^{\rho}}$ \cite{polettini14}
\begin{equation}
	T \bar{\dot{S}}_{\mathrm{i}} = \bar{\mathcal{J}}^{\varepsilon} \mathcal{A}_{\varepsilon} \, ,
	\label{ex:eprSS}
\end{equation}
which clearly emphasizes the crucial role of emergent cycles in steady state dissipation.

\subsubsection{Entropy Flow Rate}

The \emph{entropy flow rate} measures the reversible entropy changes in the environment due to exchange processes with the system \cite{kondepudi14}.
Using the expressions for the enthalpy of reaction \eqref{ex:hor} and entropy of formation \eqref{ex:cpDEC}, we express the \emph{entropy flow rate} as
\begin{equation}
	T \efr \equiv \underbrace{J^{\rho} \, \dr{H}}_{\textstyle \equiv \dot{Q}} + I^{\sigma_{\mathrm{y}}} \, T s_{\sigma_{\mathrm{y}}} \, .
	\label{ex:efrQ}
\end{equation}
The first contribution is the \emph{heat flow rate} (positive if heat is absorbed by the system). When divided by temperature, it measures minus the entropy changes in the thermal bath.
The second contribution accounts for minus the entropy change in the chemostats.

\subsubsection{System Entropy}
\label{sec:entropy}

The entropy of the ideal dilute solution constituting the CRN is given by (see app.~\ref{app:idl})
\begin{equation}
	S = Z^{\sigma} \, s_{\sigma} + R \, Z^{\mathcal{S}} + S_{0} \, .
	\label{ex:se}
\end{equation}
The total concentration term
\begin{equation}
	Z^{\mathcal{S}} \equiv \sum_{\sigma \in \mathcal{S}} Z^{\sigma}
	\label{def:totalConcentration}
\end{equation}
and the constant $S_{0}$ together represent the entropic contribution of the solvent.
$S_{0}$ may also account for the entropy of chemical species not involved in the reactions.
We also prove in app.~\ref{sec:stochEntropy} that the entropy \eqref{ex:se} can be obtained as a large particle limit of the stochastic entropy of CRNs.

$S$ would be an equilibrium entropy if the reactions could be all shut down.
But in presence of reactions, it becomes the \emph{nonequilibrium entropy} of the CRN.
Indeed, using eqs. \eqref{ex:cpDEC}, \eqref{ex:eprG} and \eqref{ex:efrQ}, we find that its change can be expressed as 
\begin{equation}
	\begin{split}
		\dt{S} & = s_{\sigma} \, \dt{Z}^{\sigma} + {Z}^{\sigma} \, \dt s_{\sigma} + R \, \dt Z^{\mathcal{S}} \\
		& = s_{\sigma} \, \dt{Z}^{\sigma} \\
		& = J^{\rho} \, \dr{S} + I^{\sigma_{\mathrm{y}}} \, s_{\sigma_{\mathrm{y}}} \\
		& = \epr + \efr \, .
	\end{split}
	\label{ex:sec}
\end{equation}
This relation is the nonequilibrium formulation of the \emph{Second Law of Thermodynamics} for CRNs.
It demonstrates that the non-negative entropy production (\ref{ex:eprJ}) measures the entropy changes in the system plus those in the reservoirs (thermal and chemostats) \cite{kondepudi14}.

\subsection{Energy Balance}
\label{sec:energyBalance}

\subsubsection{First Law of Thermodynamics}
\label{sec:Ilaw}

Since the CRN is kept at constant pressure $p$, its enthalpy 
\begin{equation}
	H = Z^{\sigma} \, h^\circ_{\sigma} + H_{0} \, 
	\label{ex:H}
\end{equation}
is equal to the CRN internal energy, up to a constant.
Indeed, the enthalpy $H$ is a density which, when written in terms of the internal energy (density) $U$, reads $H = U + p$.

Using the rate equations \eqref{ex:intflux} and \eqref{ex:extflux}, the enthalpy rate of change can be expressed as the sum of the heat flow rate, defined in eq.~\eqref{ex:efrQ}, and the enthalpy of formation exchange rate
\begin{equation}
		\dt{H} = h^\circ_{\sigma} \, \dt Z^{\sigma} = \dot{Q} + I^{\sigma_{\mathrm{y}}} \, h^\circ_{\sigma_{\mathrm{y}}} \, .
	\label{ex:Hdot}
\end{equation}
Equivalently, it can be rewritten in terms of the entropy flow rate \eqref{ex:efrQ} as \cite[\S~4.1.2]{kondepudi14}
\begin{equation}
	\dt{H} = T \efr + I^{\sigma_{\mathrm{y}}} \, \mu_{\sigma_{\mathrm{y}}} \, .
	\label{ex:efrH}
\end{equation}
The last term on the r.h.s. of eq.~\eqref{ex:efrH} is the free energy exchanged with the chemostats. 
It represents the \emph{chemical work rate} performed by the chemostats on the CRN \cite{alberty03,schmiedl07}
	\begin{equation}
		\cwr \equiv I^{\sigma_{\mathrm{y}}} \, \mu_{\sigma_{\mathrm{y}}} \, .
		\label{def:cw}
	\end{equation}
Either eq.~\eqref{ex:Hdot} or \eqref{ex:efrH} may be considered as the nonequilibrium formulation of the \emph{First Law of Thermodynamics} for CRNs.
The former has the advantage to solely focus on energy exchanges. The latter contains entropic contributions but is appealing because it involves the chemical work \eqref{def:cw}.

\subsubsection{Nonequilibrium Gibbs Free Energy}
\label{sec:CA}

We are now in the position to introduce the thermodynamic potential regulating CRNs.
The Gibbs free energy of ideal dilute solutions reads 
\begin{equation}
	G \equiv H -T S = Z^{\sigma} \, \mu_{\sigma} - RT \, Z^{\mathcal{S}} + G_{0} \, .
	\label{ex:G}
\end{equation}
As for entropy, the total concentration term $-RT \, Z^{\mathcal{S}}$ and the constant $G_{0}$ represent the contribution of the solvent (see app.~\ref{app:idl}).
Furthermore, in presence of reactions, $G$ becomes the \emph{nonequilibrium Gibbs free energy} of CRNs.

We will now show that the nonequilibrium Gibbs free energy of a closed CRN is always greater or equal than its corresponding equilibrium form.
A generic nonequilibrium concentration distribution $\set{Z^{\sigma}}$ is characterized by the set of components $\{ L^{\lambda} = \ell^{\lambda}_{\sigma} \, Z^{\sigma}\}$. Let $\{ Z_{\mathrm{eq}}^{\sigma} \}$ be the corresponding equilibrium distribution defined by the detailed-balance property \eqref{eq:db} and characterized by the same set of components $\{ L^{\lambda} \}$ (a formal expression for the equilibrium distribution will be given in eq.~\eqref{ex:equilibrium}).
At equilibrium, the Gibbs free energy \eqref{ex:G} reads
\begin{equation}
	G_{\mathrm{eq}} = Z_{\mathrm{eq}}^{\sigma} \, \mu^{\mathrm{eq}}_{\sigma} - RT \, Z^{\mathcal{S}}_{\mathrm{eq}} + G_{0} \, .
	\label{ex:Geq}
\end{equation}
As discussed in \S~\ref{sec:affinities}, the equilibrium chemical potentials must satisfy $\nabla^{\sigma}_{\rho} \, \mu^{\mathrm{eq}}_{\sigma} = 0$.
We deduce that $\mu^{\mathrm{eq}}_{\sigma}$ must be a linear combination of the closed system conservation laws \eqref{def:cl}
\begin{equation}
	\mu^{\mathrm{eq}}_{\sigma} = {f}_{\lambda} \, \ell^{\lambda}_{\sigma} \, ,
	\label{eq:cpeq}
\end{equation}
where $\set{f_{\lambda}}$ are real coefficients.
Thus, we can write the equilibrium Gibbs free energy as
\begin{equation}
	G_{\mathrm{eq}} = {f}_{\lambda} \, L^{\lambda} - RT \, Z^{\mathcal{S}}_{\mathrm{eq}} + G_{0} \, .
	\label{ex:GeqF}
\end{equation}
In this form, the first term of the Gibbs free energy appears as a bilinear form of components $\set{L^{\lambda}}$ and \emph{conjugated generalized forces} $\set{{f}_{\lambda}}$ \cite[\S~3.3]{alberty03}, which can be thought of as chemical potentials of the components.
From eq.~\eqref{eq:cpeq} and the properties of components \eqref{eq:component}, the equality $Z_{\mathrm{eq}}^{\sigma} \, \mu^{\mathrm{eq}}_{\sigma} = Z^{\sigma} \, \mu^{\mathrm{eq}}_{\sigma}$ follows.
Hence, using the definition of chemical potential \eqref{ex:cp}, the nonequilibrium Gibbs free energy $G$ of the generic distribution $\set{Z^{\sigma}}$ defined above is related to $G_{\mathrm{eq}}$ \eqref{ex:GeqF} by
\begin{equation}
	G = G_{\mathrm{eq}} + RT \Ly{Z}{Z_{\mathrm{eq}}}{} \, ,
	\label{eq:GLGeq}
\end{equation}
where we introduced the relative entropy for non-normalized concentration distributions, also called \emph{Shear Lyapunov Function}, or \emph{pseudo-Helmholtz function} \cite{shear67,higgins68,horn72}
\begin{equation}
	\hspace{-0.2cm}
	\Ly{Z}{Z'}{} \equiv Z^{\sigma} \ln \frac{Z_{\sigma}}{Z'_{\sigma}} - \left( Z^{\mathcal{S}} - Z'^{\mathcal{S}} \right) \geq 0 \, .
	\label{def:shear}
\end{equation}
This quantity is a natural generalization of the relative entropy, or Kullback--Leibler divergence, used to compare two normalized probability distributions \cite{cover06}. 
For simplicity, we still refer to it as \emph{relative entropy}.
It quantifies the distance between two distributions:
it is always positive and vanishes only if the two distributions are identical, $\left\{ Z^{\sigma} \right\} = \left\{ Z'^{\sigma} \right\}$.
Hence, eq.~\eqref{eq:GLGeq} proves that the nonequilibrium Gibbs free energy of a closed CRN is always greater or equal than its corresponding equilibrium form, $G \ge G_{\mathrm{eq}}$. 

We now proceed to show that the nonequilibrium Gibbs free energy is minimized by the dynamics in closed CRNs, \emph{viz.} $G$---or equivalently $\Ly{Z}{Z_{\mathrm{eq}}}{}$ \cite{schuster89,schaft13}---acts as a Lyapunov function in closed CRNs.
Indeed, the time derivative of $G$ \eqref{ex:G} always reads
\begin{equation}
	\begin{split}
		\dt{G} & = \mu_{\sigma} \, \dt{Z}^{\sigma} + {Z}^{\sigma} \, \dt \mu_{\sigma} + R \, \dt Z^{\mathcal{S}} \\ 
		& = \mu_{\sigma} \, \dt{Z}^{\sigma} \, .
	\end{split}
	\label{eq:SHG}
\end{equation}
When using the rate equation for closed CRNs \eqref{eq:rate} we find that $\dt{G} = - J^{\rho} \, \nabla^{\sigma}_{\rho} \, \mu_{\sigma}$.
Using eq.~\eqref{eq:GLGeq} together with eqs.~\eqref{ex:gfer} and \eqref{ex:eprG}, we get
\begin{equation}
		\dt{G} = RT \, \dt \Ly{Z}{Z_{\mathrm{eq}}}{} = - T \epr \le 0 \, ,
	\label{}
\end{equation}
which proves the aforementioned result.

\subsubsection{Chemical Work}
\label{sec:cw}

In arbitrary CRNs, the rate of change of nonequilibrium Gibbs free energy \eqref{eq:SHG} can be related to the entropy production rate \eqref{ex:erpIED} using the rate equations of open CRN \eqref{ex:intflux} and \eqref{ex:extflux} and the chemical work rate \eqref{def:cw},
	\begin{equation}
		T \epr = \cwr- \dt{G}  \geq 0 \, .
		\label{ex:eprGp}
	\end{equation}
This important results shows that the positivity of the entropy production sets an intrinsic limit on the chemical work that the chemostats must perform on the CRN to change its concentration distribution. 
The equality sign is achieved for quasi-static transformations ($\epr \simeq 0$).

\begin{figure}[t]
	\centering
	\includegraphics[width=.48\textwidth]{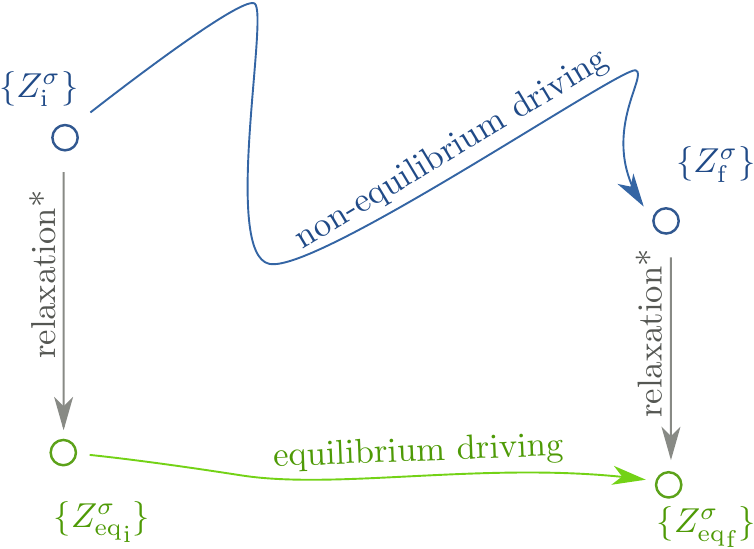}
	\caption{
		Pictorial representation of the transformation between two nonequilibrium concentration distributions.
		The nonequilibrium transformation (blue line) is compared with the equilibrium one (green line).
		The equilibrium transformation depends on the equilibrium states corresponding to the initial and final concentration distributions.
		In \S~\ref{sec:cw}, for an arbitrary CRN, these equilibrium states are obtained by first closing the network and then letting it relax to equilibrium.
		Instead, in sec.~\ref{sec:dbnThermo}, for a detailed balance CRN, the equilibrium states are obtained by simply stopping the time-dependent driving and letting the system spontaneously relax to equilibrium.
	}
	\label{fig:driving}
\end{figure}

If we now integrate eq.~\eqref{ex:eprGp} along a transformation generated by an arbitrary time dependent protocol $\pi(t)$, which drives the CRN from an initial concentration distribution $\set{Z^{\sigma}_{\mathrm{i}}}$ to a final one $\set{Z^{\sigma}_{\mathrm{f}}}$, we find
\begin{equation}
	T \Delta_{\mathrm{i}} S = W_{\mathrm{c}} - \Delta G \geq 0 \, ,
	\label{eq:W=DG+DS}
\end{equation}
where $\Delta G = G_{\mathrm{f}} - G_{\mathrm{i}}$ is the difference of nonequilibrium Gibbs free energies between the final and the initial state.
Let us also consider the equilibrium state $\set{{Z_{\mathrm{eq}}^{\sigma}}_{\mathrm{i}}}$ (resp. $\set{{Z_{\mathrm{eq}}^{\sigma}}_{\mathrm{f}}}$) obtained from $\set{Z^{\sigma}_{\mathrm{i}}}$ (resp. $\set{Z^{\sigma}_{\mathrm{f}}}$) if one closes the network (i.e. interrupt the chemostatting procedure) and let it relax to equilibrium, as illustrated in fig.~\ref{fig:driving}.
The Gibbs free energy difference between these two equilibrium distributions, $\Delta G_{\mathrm{eq}} = {G_{\mathrm{eq}}}_{\mathrm{f}} - {G_{\mathrm{eq}}}_{\mathrm{i}}$, is related to $\Delta G$ via the difference of relative entropies, eq.~\eqref{eq:GLGeq},
\begin{equation}
		\Delta G = \Delta G_{\mathrm{eq}} + RT \, \Delta \mathcal{L} \, ,
	\label{}
\end{equation}
where
\begin{equation}
	\Delta \mathcal{L} \equiv \mathcal{L}\big( \{ Z_{\mathrm{f}}^{\sigma} \} | \{ {Z_{\mathrm{eq}}^{\sigma}}_{\mathrm{f}} \} \big) - \mathcal{L}\big( \{ Z_{\mathrm{i}}^{\sigma} \} | \{ {Z_{\mathrm{eq}}^{\sigma}}_{\mathrm{i}} \} \big) \, .
	\label{ex:Dshear}
\end{equation}
Thus, the chemical work \eqref{eq:W=DG+DS} can be rewritten as
\begin{equation}
	W_{\mathrm{c}} - \Delta G_{\mathrm{eq}} = RT \Delta \mathcal{L} + T \Delta_{\mathrm{i}} S \, ,
	\label{ex:Wirr}
\end{equation}
which is a key result of our paper.
$\Delta G_{\mathrm{eq}}$ represents the reversible work needed to reversibly transform the CRN from $\set{{Z_{\mathrm{eq}}^{\sigma}}_{\mathrm{i}}}$ to $\set{{Z_{\mathrm{eq}}^{\sigma}}_{\mathrm{f}}}$.
Implementing such a reversible transformation may be difficult to achieve in practice.
However, it allows us to interpret the difference $W^{\mathrm{irr}}_{\mathrm{c}} \equiv W_{\mathrm{c}} - \Delta G_{\mathrm{eq}}$ in eq.~\eqref{ex:Wirr} as the chemical work dissipated during the nonequilibrium transformation, \emph{i.e.} the \emph{irreversible chemical work}.
The positivity of the entropy production implies that 
\begin{equation}
	W^{\mathrm{irr}}_{\mathrm{c}} \ge RT \, \Delta \mathcal{L} \, .
	\label{ineq:Wirr}
\end{equation}
This relation sets limits on the irreversible chemical work involved in arbitrary far-from-equilibrium transformations. 
For transformations connecting two equilibrium distributions, we get the expected inequality $W^{\mathrm{irr}}_{\mathrm{c}} \ge 0$.
More interestingly, eq.~\eqref{ineq:Wirr} tells us how much chemical work the chemostat need to provide to create a nonequilibrium distribution from an equilibrium one.
It also tells us how much chemical work can be extracted from a CRN relaxing to equilibrium.

The conceptual analogue of (\ref{ex:Wirr}) in Stochastic Thermodynamics (where probability distributions replace non-normalized concentration distributions) is called the nonequilibrium Landauer's principle \cite{takara10,esposito11} (see also \cite{procaccia76,vaikuntanathan09, hasegawa10}).
It has been shown to play a crucial role to analyze the thermodynamic cost of information processing (\emph{e.g.} for Maxwell demons, feedback control or proofreading). The inequality \eqref{ineq:Wirr} is therefore a \emph{nonequilibrium Landauer's principle} for CRN.

\section{Thermodynamics of Complex-Balanced Networks}
\label{sec:cbnThermo}

In this section, we focus on unconditionally complex-balanced networks.
We shall see that the thermodynamics of these networks bears remarkable similarities with Stochastic Thermodynamics.

Let us first observe that whenever a CRN displays a well-defined steady-state distribution $\set{\bar{Z}^{\sigma_{\mathrm{x}}}}$, 
the entropy production rate \eqref{ex:eprJ} can be formally decomposed as the sum of an \emph{adiabatic} and a \emph{nonadiabatic} contribution
	\begin{equation}
		T \epr = \underbrace{ J^{\rho} \, RT \, \ln \frac{\bar{J}_{+\rho}}{\bar{J}_{-\rho}} }_{\textstyle \equiv T \vepr{a}} \underbrace{ - \dt{Z}^{\sigma_{\mathrm{x}}} \, RT \, \ln \frac{Z_{\sigma_{\mathrm{x}}}}{\bar{Z}_{\sigma_{\mathrm{x}}}} }_{\textstyle \equiv T \vepr{na}}\, ,
		\label{ex:ANAdec}
	\end{equation}
in analogy to what was done in Stochastic Thermodynamics \cite{esposito07, harris07, esposito10:threefaces1, vandenbroeck10, ge10}.
As discussed in \S~\ref{sec:cbn}, unconditionally complex-balanced networks have a unique steady-state distribution $\left\{ \bar{Z}^{\sigma_{\mathrm{x}}} \equiv \bar{Z}^{\sigma_{\mathrm{x}}}(\pi(t)) \right\}$, eq.~\eqref{def:cb}, for any value of the chemostatted concentrations $\{ Z^{\sigma_{\mathrm{y}}} \equiv Z^{\sigma_{\mathrm{y}}}(\pi(t)) \}$ and of the fixed unbroken components $\set{L^{\lambda_{\mathrm{u}}}}$.
The decomposition \eqref{ex:ANAdec} is thus well-defined at any time, for any protocol $\pi(t)$.
As a central result, we prove in app.~\ref{sec:anadec} that the adiabatic and nonadiabatic contribution are non-negative for unconditionally complex-balanced networks as well as for complex-balanced networks without time-dependent driving.

The adiabatic entropy production rate encodes the dissipation of the steady state $\left\{ \bar{Z}^{\sigma_{\mathrm{x}}} \right\}$.
It can be rewritten in terms of the steady state Gibbs free energy of reaction $\set{\dr{\bar{G}}}$ \eqref{ex:gferJ} as
\begin{equation}
	T \vepr{a} = - J^{\rho} \, \Delta_{\mathrm{r}} \bar{G}_{\rho} \ge 0 \, .
	\label{}
\end{equation}
This inequality highlights the fact that the transient dynamics---generating the currents $\set{J^{\rho}}$---is constrained by the thermodynamics of the complex balanced steady state, \emph{i.e.} by $\set{\dr{\bar{G}}}$.

The nonadiabatic entropy production rate characterizes the dissipation of the transient dynamics.
It can be decomposed as
\begin{multline}
	T \vepr{na} = - RT \, \dt \LyX{Z}{\bar{Z}} + \\
	+ \underbrace{ RT \, \dt {\bar{Z}^{\mathcal{S}_{\mathrm{x}}}} - Z^{\sigma_{\mathrm{x}}} \dt {\bar{\mu}_{\sigma_{\mathrm{x}}}} }_{\textstyle \equiv T \vepr{d}} \ge 0 \, ,
	\label{ex:NAepr}
\end{multline}
where ${\bar{Z}^{\mathcal{S}_{\mathrm{x}}}} = \sum_{\sigma_{\mathrm{x}} \in \mathcal{S}_{\mathrm{x}}} \bar{Z}^{\sigma_{\mathrm{x}}}$ (see Refs.~\cite{esposito07,esposito10:threefaces1} for the analogous decomposition in the stochastic context).
The first term is proportional to the time derivative of the relative entropy \eqref{def:shear} between the nonequilibrium concentration distribution at time $t$ and the corresponding complex-balanced steady-state distribution.
Hence, it describes the dissipation of the relaxation towards the steady state.
The second term, $T \dot{S}_{\mathrm{d}}$, is related to the time-dependent driving performed via the chemostatted species and thus denoted \emph{driving entropy production rate} \cite{esposito07}.
It vanishes in nondriven networks where we obtain
\begin{equation}
	\vepr{na} = - R \, \dt \LyX{Z}{\bar{Z}} \ge 0 \, .
	\label{ex:NAerpND}
\end{equation}
This result shows the role of the relative entropy $\LyX{Z}{\bar{Z}}$ as a Lyapunov function in nondriven complex-balanced networks with mass action kinetics.
It was known in the mathematical literature \cite{horn72,gopalkrishnan13}, but we provide a clear thermodynamic interpretation to this result by demonstrating that it derives from the nonadiabatic entropy production rate.

We mention that an alternative derivation of the adiabatic--nonadiabatic decomposition for nondriven complex-balanced networks with mass action kinetics was found in Ref.~\cite{ge16}, while we were finalizing our paper.

\section{Thermodynamics of Open Detailed-Balanced Networks}
\label{sec:dbnThermo}

We finish our study by considering detailed-balanced networks. 
We discuss the equilibrium distribution, introduce a new class of nonequilibrium potentials and derive a new work inequality.

Let us also emphasize that open detailed-balanced CRNs are a special class of open complex-balanced CRNs for which the adiabatic entropy production rate vanishes (since the steady state is detailed balanced) and thus the nonadiabatic entropy production characterizes the entire dissipation.

\subsection{Equilibrium Distribution}
\label{sec:ed}

As discussed in \S~\ref{sec:dbn}, for given kinetics $\set{k^{\pm \rho}}$, chemostatting $\set{Z^{\sigma_{\mathrm{y}}}}$ and unbroken components $\set{L^{\lambda_{\mathrm{u}}}}$, detailed-balanced networks always relax to a unique equilibrium distribution.
Since the equilibrium chemical potentials can be expressed as a linear combination of conservation laws, eq.~\eqref{eq:cpeq}, we can express the equilibrium distribution as 
\begin{equation}
	Z^{\mathrm{eq}}_{\sigma} = \exp \left\{ - \frac{\mu^\circ_{\sigma} - {f}_{\lambda} \, \ell^{\lambda}_{\sigma}}{RT} \right\} \, ,
	\label{ex:equilibrium}
\end{equation}
inverting the expression for the chemical potentials \eqref{ex:cp}.
Since the independent set of unbroken conservation laws, $\set{\bm\ell^{\lambda_{\mathrm{u}}}}$, are such that $\ell^{\lambda_{u}}_{\sigma_{\mathrm{y}}} = 0, \, \forall \lambda_{\mathrm{u}},\sigma_{\mathrm{y}}$, see \S~\ref{sec:cl}, we have that
\begin{equation}
	\mu^{\mathrm{eq}}_{\sigma_{\mathrm{y}}} = {f}_{\lambda_{\mathrm{b}}} \, \ell^{\lambda_{\mathrm{b}}}_{\sigma_{\mathrm{y}}} \, , \quad \forall \, \sigma_{\mathrm{y}} \in \mathcal{S}_{\mathrm{y}} \, .
	\label{eq:muy=fly}
\end{equation}
We thus conclude that the $|\lambda_{\mathrm{b}}|$ broken generalized forces $\set{f_{\lambda_{\mathrm{b}}}}$ only depend on the chemostatted concentrations $\set{Z^{\sigma_{\mathrm{y}}}}$.
Instead, the remaining $|\lambda_{\mathrm{u}}|$ unbroken generalized forces $f_{\lambda_{\mathrm{u}}}$ can be determined by inverting the nonlinear set of equations $L^{\lambda_{\mathrm{u}}} = \ell^{\lambda_{\mathrm{u}}}_{\sigma_{\mathrm{x}}} Z^{\sigma_{\mathrm{x}}}_{\mathrm{eq}}$.
They therefore depend on both $\set{Z^{\sigma_{\mathrm{y}}}}$ and $\set{L^{\lambda_{\mathrm{u}}}}$.

One can easily recover the local detailed-balanced property \eqref{ex:ldb} and \eqref{eq:db} using eq.~\eqref{ex:equilibrium}.

\subsection{Open nondriven networks}
\label{sec:tgfe}

As a consequence of the break of conservation laws, the nonequilibrium Gibbs free energy $G$ \eqref{ex:G} is no longer minimized at equilibrium in open detailed-balanced networks.
In analogy to equilibrium thermodynamics \cite{alberty03}, the proper thermodynamic potential is obtained from $G$ by subtracting the energetic contribution of the broken conservation laws.
This \emph{transformed nonequilibrium Gibbs free energy} reads
\begin{equation}
	\begin{split}
		\mathcal{G} & \equiv G - {f}_{\lambda_{\mathrm{b}}} \, L^{\lambda_{\mathrm{b}}} \\
		& = Z^{\sigma} \left( \mu_{\sigma} - {f}_{\lambda_{\mathrm{b}}} \, \ell^{\lambda_{\mathrm{b}}}_{\sigma} \right) - RT \, Z^{\mathcal{S}} + G_{0} \, .
	\end{split}
	\label{def:tca}
\end{equation}

We proceed to show that $\mathcal{G}$ is minimized by the dynamics in nondriven open detailed-balanced networks.
Let $\{ Z^{\sigma_{\mathrm{x}}} \}$ be a generic concentration distribution in a detailed-balanced network characterized by $\set{L^{\lambda_{\mathrm{u}}}}$ and $\set{Z^{\sigma_{\mathrm{y}}}}$, and let $\set{Z_{\mathrm{eq}}^{\sigma_{\mathrm{x}}}}$ be its corresponding equilibrium.
Using the relation between equilibrium chemical potentials and conservation laws \eqref{eq:cpeq}, the transformed Gibbs free energy \eqref{def:tca} at equilibrium reads
\begin{equation}
	\mathcal{G}_{\mathrm{eq}} = {f}_{\lambda_{\mathrm{u}}} \, L^{\lambda_{\mathrm{u}}} - RT \, Z^{\mathcal{S}}_{\mathrm{eq}} + G_{0} \, .
	\label{ex:tGeq}
\end{equation}
Yet, combinig eq.~\eqref{eq:cpeq} and the properties of unbroken components, one can readily show that $Z_{\mathrm{eq}}^{\sigma} \left( \mu^{\mathrm{eq}}_{\sigma} - {f}_{\lambda_{\mathrm{b}}} \, \ell^{\lambda_{\mathrm{b}}}_{\sigma} \right) = Z^{\sigma} \left( \mu^{\mathrm{eq}}_{\sigma} - {f}_{\lambda_{\mathrm{b}}} \, \ell^{\lambda_{\mathrm{b}}}_{\sigma} \right)$.
The relation between the nonequilibrium $\mathcal{G}$ and the corresponding equilibrium value thus follows
\begin{equation}
	\mathcal{G} = \mathcal{G}_{\mathrm{eq}} + RT \LyX{Z}{Z_{\mathrm{eq}}}
	\label{eq:tGLtGeq}
\end{equation}
(we show in app.~\ref{app:idl} the derivation of the latter in presence of reacting solvent).
The non-negativity of the relative entropy for concentration distributions $\LyX{Z}{Z_{\mathrm{eq}}}$ ensures that the nonequilibrium transformed Gibbs free energy is always greater or equal to its equilibrium value, $\mathcal{G} \ge \mathcal{G}_{\mathrm{eq}}$.
Since entropy production and nonadiabatic entropy production coincide, using eqs.~\eqref{ex:NAerpND} and \eqref{eq:tGLtGeq}, we obtain
\begin{equation}
	\dt{\mathcal{G}} = RT \, \dt \LyX{Z}{Z_{\mathrm{eq}}} = - T \epr \leq 0 \, ,
	\label{eq:dotGG=-epr}
\end{equation}
which demonstrates the role of $\mathcal{G}$ as a Lyapunov function.
The relative entropy $\LyX{Z}{Z'}$ was known to be a Lyapunov function for detailed-balanced networks \cite{schaft13,schaft15}, but we provided its clear connection to the transformed nonequilibrium Gibbs free energy.
To summarize, instead of minimizing the nonequilibrium Gibbs free energy $G$ \eqref{ex:G} as in closed CRNs, the dynamics minimizes the transformed nonequilibrium Gibbs free energy $\mathcal{G}$ in open nondriven detailed-balanced CRNs.

\subsection{Open driven networks}
\label{sec:udbnThermo}

We now consider unconditionally detailed-balanced CRNs.
As discussed in \S~\ref{sec:dbn}, they are characterized by a unique equilibrium distribution $\left\{ Z_{\mathrm{eq}}^{\sigma_{\mathrm{x}}} \equiv Z_{\mathrm{eq}}^{\sigma_{\mathrm{x}}}(\pi(t)) \right\}$, defined by eq.~\eqref{eq:db}, for any value of the chemostatted concentrations $\{ Z^{\sigma_{\mathrm{y}}} = Z^{\sigma_{\mathrm{y}}}(\pi(t)) \}$.

We start by showing that the external fluxes $\set{I^{\sigma_{\mathrm{y}}}}$ can be expressed as influx rate of moieties.
Since the CRN is open and unconditionally detailed balanced, each chemostatted species broke a conservation law (no emergent cycle is created, \S~\ref{sec:dbn}).
Therefore, the matrix whose entries are $\set{\ell^{\lambda_{\mathrm{b}}}_{\sigma_{\mathrm{y}}}}$ in eq.~\eqref{eq:muy=fly} is square and also nonsingular 
\footnote{The fact that the matrix whose entries are $\set{\ell^{\lambda_{\mathrm{b}}}_{\sigma_{\mathrm{y}}}}$ is nonsingular follows from $\bm\ell^{\lambda_{\mathrm{u}}} \in \ker \nabla\, , \forall \, \lambda_{\mathrm{u}}$, and from the linear independence of $\set{\bm\ell^{\lambda}}$.}.
We can thus invert eq.~\eqref{eq:muy=fly} to get
\begin{equation}
	f_{\lambda_{\mathrm{b}}} = \mu^{\mathrm{eq}}_{\sigma_{\mathrm{y}}} \, \hat{\ell}^{\sigma_{\mathrm{y}}}_{\lambda_{\mathrm{b}}} \, ,
	\label{eq:f=mul}
\end{equation}
where $\set{\hat{\ell}^{\sigma_{\mathrm{y}}}_{\lambda_{\mathrm{b}}}}$ denote the entries of the inverse matrix of that with entries $\set{\ell_{\sigma_{\mathrm{y}}}^{\lambda_{\mathrm{b}}}}$.
Hence, using the definition of broken component, $\set{L^{\lambda_{\mathrm{b}}} \equiv \ell^{\lambda_{\mathrm{b}}}_{\sigma} Z^{\sigma}}$, we obtain that
\begin{equation}
	f_{\lambda_{\mathrm{b}}} \, L^{\lambda_{\mathrm{b}}} = \mu^{\mathrm{eq}}_{\sigma_{\mathrm{y}}} \, \underbrace{\hat{\ell}^{\sigma_{\mathrm{y}}}_{\lambda_{\mathrm{b}}} \, \ell^{\lambda_{\mathrm{b}}}_{\sigma} \, Z^{\sigma}}_{\textstyle \equiv M^{\sigma_{\mathrm{y}}}} \, .
	\label{eq:fL=muM}
\end{equation}
From the rate equations for the chemostatted concentrations \eqref{ex:extflux}, we find that
\begin{equation}
	\dt{M}^{\sigma_{\mathrm{y}}} = I^{\sigma_{\mathrm{y}}} \, , \quad \forall \, \sigma_{\mathrm{y}} \in \mathcal{S}_{\mathrm{y}} \, .
	\label{eq:Mdot=I}
\end{equation}
We can thus interpret $M^{\sigma_{\mathrm{y}}}$ as the concentration of a moiety which is exchanged with the environment only through the chemostatted species $\ce{X}_{\sigma_{\mathrm{y}}}$.
Eq.~\eqref{eq:fL=muM} shows that the energetic contribution of the broken components can be expressed as the Gibbs free energy carried by these specific moieties.

\begin{example}
	A simple implementation of this scenario is the thermodynamic description of CRNs at constant pH \cite[Ch.~4]{alberty03} where the chemostatted species becomes the ion $\mathrm{H}^{+}$ and ${M}^{\mathrm{H}^{+}}$ is the total amount of $\mathrm{H}^{+}$ ions in the system.
	The transformed Gibbs potential thus become $G' = G - \mu_{\mathrm{H}_{+}} {M}^{\mathrm{H}^{+}}$ and the transformed chemical potentials can be written in our formalism as $\mu'_{\sigma_{\mathrm{x}}} = \mu_{\sigma_{\mathrm{x}}} - \mu_{\mathrm{H}_{+}} \hat{\ell}_{\mathrm{b}}^{ \, \mathrm{H}_{+}} \, \ell^{\mathrm{b}}_{\sigma}$, where $\ell^{\mathrm{b}}_{\sigma}$ is the conservation law broken by chemostatting $\mathrm{H}^{+}$.
	\qed
\end{example}

\begin{example}
	For the CRN in fig.~\ref{fig:oCRN}, whose conservation laws given in example~\ref{exm:cl}, the concentrations of the exchanged moieties are
	\begin{equation}
		\begin{aligned}
			M^{1} & = Z^{\mathrm{a}} + \tfrac{1}{2} Z^{\mathrm{b}} \\
			M^{2} & = Z^{\mathrm{d}} + Z^{\mathrm{e}} \, .
		\end{aligned}
		\label{}
	\end{equation}
	For the specific implementation of that CRN, fig.~\ref{fig:eCRN}, the first term (resp. second term) is the total number of moiety $2\mathrm{H}$ (resp. $\mathrm{C}$) in the system, which can be exchanged with the environment only via the chemostatted species $\ce{H_{2}O}$ (resp. $\ce{CO}$). 
	\qed
\end{example}

We now turn to the new work relation. 
From the general work relation \eqref{ex:eprGp}, using \eqref{def:tca} and \eqref{eq:fL=muM}, we find
\begin{equation}
	T \epr = \dwr - \dt{\mathcal{G}} \geq 0 \, ,
	\label{eq:epr=G+Wdirr}
\end{equation}
where the driving work due to the time-dependent driving of the chemostatted species is obtained using the chemical work rate \eqref{def:cw} together with eqs.~\eqref{eq:fL=muM} and \eqref{eq:Mdot=I}
\begin{equation}
	\begin{split}
		\dwr & \equiv \cwr - \dt\left( f_{\lambda_{\mathrm{b}}} \, L^{\lambda_{\mathrm{b}}} \right) \\
		& = {\mu}^{\mathrm{eq}}_{\sigma_{\mathrm{y}}} \, \dt M^{\sigma_{\mathrm{y}}} - \dt \big( {\mu}^{\mathrm{eq}}_{\sigma_{\mathrm{y}}} \, M^{\sigma_{\mathrm{y}}} \big) \\
		& = - \dt{\mu}^{\mathrm{eq}}_{\sigma_{\mathrm{y}}} \, M^{\sigma_{\mathrm{y}}} \, .
	\end{split}
	\label{eq:epr=G+WdirrBis}
\end{equation}
Equivalently, the driving work rate \eqref{eq:epr=G+WdirrBis} can be defined as the rate of change of the transformed Gibbs free energy \eqref{def:tca} due to the time dependent driving only, \emph{i.e.}
\begin{equation}
	\dot{W}_{\mathrm{d}} \equiv \derpart{\mathcal{G}}{t} \equiv \dt \mu^{\mathrm{eq}}_{\sigma_{\mathrm{y}}} \, \derpart{\mathcal{G}}{\mu^{\mathrm{eq}}_{\sigma_{\mathrm{y}}}} \, .
	\label{}
\end{equation}
To relate this alternative definition to eq.~\eqref{eq:epr=G+WdirrBis}, all $\set{Z^{\sigma_{\mathrm{y}}}}$ must be expressed in terms of $\set{\mu^{\mathrm{eq}}_{\sigma_{\mathrm{y}}}}$ using the definition of chemical potential \eqref{ex:cp}.

The driving work rate $\dot{W}_{\mathrm{d}}$ vanishes in nondriven CRNs, where \eqref{eq:epr=G+Wdirr} reduces to \eqref{eq:dotGG=-epr}.
After demonstrating that the entropy production rate is always proportional the difference between the chemical work rate and the change of nonequilibrium Gibbs free energy in eq.~\eqref{eq:W=DG+DS}, we showed that, for unconditionally detailed-balanced CRNs, it is also proportional the difference between the driving work rate and the change in transformed nonequilibrium Gibbs free energy, eq.~\eqref{eq:epr=G+Wdirr}.

We end by formulating a nonequilibrium Landauer's principle for the driving work instead the chemical work done in \S~\ref{sec:cw}. 
We consider a time-dependent transformation driving the unconditionally detailed-balanced CRN from $\set{Z^{\sigma}_{\mathrm{i}}}$ to $\set{Z^{\sigma}_{\mathrm{f}}}$.
The distribution $\set{{Z_{\mathrm{eq}}^{\sigma}}_{\mathrm{i}}}$ (resp. $\set{{Z_{\mathrm{eq}}^{\sigma}}_{\mathrm{f}}}$) denotes the equilibrium distribution obtained from $\set{Z^{\sigma}_{\mathrm{i}}}$ (resp. $\set{Z^{\sigma}_{\mathrm{f}}}$) by stopping the time-dependent driving and letting the system relax towards the equilibrium, fig.~\ref{fig:driving}. Note that this reference equilibrium state is different from the one obtained by closing the network in \S~\ref{sec:cw}.
Integrating \eqref{eq:epr=G+Wdirr} over time and using \eqref{eq:tGLtGeq}, we get
\begin{equation}
	W_{\mathrm{d}} - \Delta \mathcal{G}_{\mathrm{eq}} = RT \Delta \mathcal{L} + T \Delta_{\mathrm{i}} S \, ,
	\label{ex:Wdirr}
\end{equation}
where
\begin{equation}
	\Delta \mathcal{L} \equiv \mathcal{L}\big( \{ Z_{\mathrm{f}}^{\sigma_{\mathrm{x}}} \} | \{ {Z_{\mathrm{eq}}^{\sigma_{\mathrm{x}}}}_{\mathrm{f}} \} \big) - \mathcal{L}\big( \{ Z_{\mathrm{i}}^{\sigma_{\mathrm{x}}} \} | \{ {Z_{\mathrm{eq}}^{\sigma_{\mathrm{x}}}}_{\mathrm{i}} \} \big) \, .
	\label{ex:Dshear2}
\end{equation}
$\Delta \mathcal{G}_{\mathrm{eq}}$ represents the reversible driving work, and the irreversible driving work satisfies the inequality
\begin{equation}
	W_{\mathrm{d}}^{\mathrm{irr}} \equiv W_{\mathrm{d}} - \Delta \mathcal{G}_{\mathrm{eq}} \ge RT \Delta \mathcal{L} \, .
	\label{NonEqLandDB}
\end{equation}
This central relation sets limits on the irreversible work spent to manipulate nonequilibrium distributions. 
It is a nonequilibrium Landauer's principle for the driving work by the same reasons why inequality \eqref{ineq:Wirr} is a nonequilibrium Landauer's principle for the chemical work.
The key difference is that the choice of the reference equilibrium state is different in the two cases.
The above discussed inequality (\ref{NonEqLandDB}) only holds for unconditionally detailed-balanced CRNs while eq.~\eqref{ineq:Wirr} is valid for any CRNs.

\section{Conclusions and Perspectives}
\label{sec:conclusions}

Following a strategy reminiscent of Stochastic Thermodynamics, we systematically build a nonequilibrium thermodynamic description for open driven CRNs made of elementary reactions in homogeneous ideal dilute solutions. 
The dynamics is described by deterministic rate equations whose kinetics satisfies mass action law.
Our framework in not restricted to steady states and allows to treat transients as well as time-dependent drivings performed by externally controlled chemostatted concentrations.
Our theory embeds the nonequilibrium thermodynamic framework of irreversible processes established by the Brussels School of Thermodynamics.

We now summarize our results.
Starting from the expression for the entropy production rate, we established a nonequilibrium formulation of the first and second law of thermodynamics for CRNs.
The resulting expression for the system entropy is that of an ideal dilute solution. 
The clear separation between chemostatted and internal species allowed us to identify the chemical work done by the chemostats on the CRN and to relate it to the nonequilibrium Gibbs potential.
We were also able to express the minimal chemical work necessary to change the nonequilibrium distribution of species in the CRN as a difference of relative entropies for non-normalized distributions.
These latter measure the distance of the initial and final concentration distributions from their corresponding equilibrium ones, obtained by closing the network.
This result is reminiscent of the nonequilibrium Landauer's principle derived in Stochastic Thermodynamics \cite{esposito11} and which proved very useful to study the energetic cost of information processing \cite{parrondo15}.
We also highlighted the deep relationship between the topology of CRNs, their dynamics and their thermodynamics. 
Closed CRNs (resp. nondriven open detailed-balanced networks) always relax to a unique equilibrium by minimizing their nonequilibrium Gibbs free energy (resp. transformed nonequilibrium Gibbs free energy).
This latter is given, up to a constant, by the relative entropy between the nonequilibrium and equilibrium concentration distribution.
Non-driven complex-balanced networks relax to complex-balanced nonequilibrium steady states by minimizing the relative entropy between the nonequilibrium and steady state concentration distribution. 
In all these cases, even in presence of driving, we showed how the rate of change of the relative entropy relates to the CRN dissipation.
For complex-balanced networks, we also demonstrated that the entropy production rate can be decomposed, as in Stochastic Thermodynamics, in its adiabatic and nonadiabatic contributions quantifying respectively the dissipation of the steady state and of the transient dynamics.

Our framework could be used to shed new light on a broad range of problems.
We mention only a few.

Stochastic thermodynamics has been successfully used to study the thermodynamics cost of information processing in various synthetic and biological systems \cite{esposito12:demon,sagawa13,diana13,horowitz14,barato14:three,bo15}. 
However, most of these are modeled as few state systems or linear networks \cite{hill77,schnakenberg76}---\emph{e.g.} quantum dots \cite{strasberg13}, molecular motors \cite{golubeva12,altaner15} and single enzyme mechanisms \cite{seifert11,rao15:proofreading}---while biochemical networks involve more complex descriptions.
The present work overcomes this limitation. It could be used to study biological information-handling processes such as kinetic proofreading \cite{hopfield74,ninio75,hopfield80,sontag01,ge12,murugan12,murugan14} or enzyme-assisted copolymerization \cite{bennett79,andrieux08:copolymerization,andrieux09,arias-gonzalez12,sartori13,sartori15, rao15:proofreading} which have currently only been studied as single enzyme mechanisms.

Our theory could also be used to study metabolic networks.
However, these require some care, since complex enzymatic reaction mechanisms are involved \cite{cornish12}.
Nevertheless our framework provides a basis to build effective coarse-graining procedures for enzymatic reactions \cite{esposito12}.
For instance, proofreading mechanisms operating in metabolic processes could be considered \cite{linster13}.
We foresee an increasing use of thermodynamics to improve the modeling of metabolic networks, as recently shown in Refs. \cite{beard02,beard04,chakrabarti13}.

Since our framework accounts for time-dependent drivings and transient dynamics, it could be used to represent the transmission of signals through CRNs or their response to external modulations in the environment.
These features become crucial when considering problems such as signal transduction and biochemical switches \cite{qian05:switch,beard08,qian12}, biochemical oscillations \cite{goldbeter96,vilar02}, growth and self-organization in evolving bio-systems \cite{nicolis77,feistel11}, or sensory  mechanisms \cite{mehta12,sartori14,horowitz14,bo15,hartich15,wang15}.
Also, since transient signals in CRNs can be used for computation \cite{mp10,*mp11,*mp12,*mp13,*mp14,*mp15} and have been shown to display Turing-universality \cite{hjelmfelt91,hjelmfelt92,magnasco97,soloveichik08}, one should be able to study the thermodynamic cost of chemical computing \cite{bennett82}.

Finally, one could use our framework to study any process that can be described as nucleation or reversible polymerization \cite{knowles07,knowles09,cohen13,budrikis14,kartal11,lahiri15} (see also Ref.~\cite[ch.~5--6]{krapivsky10}) since these processes can be described as CRNs \cite{rao15:denzymes}.

As closing words, we believe that our results constitute an important contribution to the theoretical study of CRNs.
It does for nonlinear chemical kinetics what Stochastic Thermodynamics has done for stochastic dynamics, namely build a systematic nonequilibrium thermodynamics on top of the dynamics.
It also opens many new perspectives and builds bridges between approaches from different communities studying CRNs:
mathematicians who study CRNs as dynamical systems, physicists who study them as nonequilibrium complex systems, and biochemists as well as bioengineers who aim for accurate models of metabolic networks.

\begin{acknowledgments}
The present project was supported by the National Research Fund, Luxembourg (project FNR/A11/02 and AFR PhD Grant 2014-2, No.~9114110) as well as by the European Research Council (project 681456). 
\end{acknowledgments}

\appendix

\section{Thermodynamics of Ideal Dilute Solutions}
\label{app:idl}

We show that the nonequilibrium Gibbs free energy \eqref{ex:G} is the Gibbs free energy of an ideal dilute solution \cite[Ch.~7]{fermi56} (see also \cite{ge16}).
We also show that in open detailed-balanced networks in which the solvent reacts with the solutes, the expression of the transformed Gibbs free energy \eqref{eq:tGLtGeq} is recovered by treating the solvent as a special chemostatted species.

The Gibbs free energy (density) of an ideal dilute mixture of chemical compounds kept at constant temperature and pressure reads
\begin{equation}
	G = Z^{\sigma} \mu_{\sigma} + Z_{0} \mu_{0} \, ,
	\label{ex:Gappendix}
\end{equation}
where the labels $\sigma\in \mathcal{S}$ refer to the solutes and $0$ to the solvent.
The chemical potentials of each species \eqref{ex:cp} read
\begin{equation}
	\begin{split}
		\mu_{\sigma} & = \mu_{\sigma}^{\circ} + RT \ln \frac{Z_{\sigma}}{Z_{\mathrm{tot}}} \, , \quad \forall \, \sigma \in \mathcal{S} \\ 
		\mu_{0} & = \mu_{0}^{\circ} + RT \ln \frac{Z_{0}}{Z_{\mathrm{tot}}} \, .
	\end{split}
	\label{}
\end{equation}
Since the solution is dilute, $Z_{\mathrm{tot}} = \sum_{\sigma \in \mathcal{S}} Z^{\sigma} + Z_{0} \simeq Z_{0}$ and the standard state chemical potentials $\set{\mu^\circ_{\sigma}}$ depend on the nature of the solvent.
Hence, the chemical potentials of the solutes read
\begin{equation}
	\mu_{\sigma} \simeq \mu_{\sigma}^{\circ} + RT \ln \frac{Z_{\sigma}}{Z_{0}} \, , \quad \forall \, \sigma \in \mathcal{S} \, ,
	\label{ex:cpSolutes}
\end{equation}
while that of the solvent
\begin{equation}
	\mu_{0} \simeq \mu_{0}^\circ - RT \frac{Z^{\mathcal{S}}}{Z_{0}} \, ,
	\label{ex:cpSolvent}
\end{equation}
where $Z^{\mathcal{S}} \equiv \sum_{\sigma \in \mathcal{S}} Z^{\sigma}$.
Therefore, the Gibbs free energy \eqref{ex:Gappendix} reads
\begin{equation}
	G \simeq Z^{\sigma} \, \mu_{\sigma} + Z^{0} \, \mu^\circ_{0} - RT \, Z^{\mathcal{S}} \, ,
	\label{}
\end{equation}
which is eq.~\eqref{ex:G} in the main text, where $G_{0}$ is equal to $Z^{0} \, \mu^\circ_{0}$ plus possibly the Gibbs free energy of solutes which do not react.

We now consider the case where the solvent reacts with the solutes.
We assume that both the solutes and the solvent react according to the stoichiometric matrix
\begin{equation}
	\nabla =
	\begin{pmatrix}
		\nabla^{0} \\
		\nabla^{\mathrm{X}} \\
		\nabla^{\mathrm{Y}}
	\end{pmatrix}
	\, ,
	\label{ex:nablaSolvent}
\end{equation}
where the first row refers to the solvent, the second block of rows to the internal species and the last one to the chemostatted species.
The solvent is treated as a chemostatted species such that $\dt{Z}_{0} = 0$.

In order to recover the expression for the transformed Gibbs free energy \eqref{eq:tGLtGeq} in unconditionally detailed balanced networks, we observe that, at equilibrium
\begin{equation}
	\nabla^{\sigma}_{\rho} \, \mu_{\sigma}^{\mathrm{eq}} + \nabla^{0}_{\rho} \, \mu_{0}^{\mathrm{eq}} = 0 \, .
	\label{}
\end{equation}
Therefore, the equilibrium chemical potentials are a linear combination of the conservation laws of $\nabla$ \eqref{ex:nablaSolvent}
\begin{equation}
	\begin{aligned}
		\mu^{\mathrm{eq}}_{\sigma} & = f_{\lambda} \, \ell^{\lambda}_{\sigma} \\
		\mu^{\mathrm{eq}}_{0} & = f_{\lambda} \, \ell^{\lambda}_{0} \, .
	\end{aligned}
	\label{eq:mueqSolvent}
\end{equation}
As mentioned in the main text, \S~\ref{sec:cl}, the chemostatting procedure breaks some conservation laws, which are labeled by $\lambda_{\mathrm{b}}$.
The unbroken ones are labeled by $\lambda_{\mathrm{u}}$.

The transformed Gibbs free energy is defined as in eq.~\eqref{def:tca}, reported here for convenience
\begin{equation}
	\mathcal{G} \equiv G - {f}_{\lambda_{\mathrm{b}}} \, L^{\lambda_{\mathrm{b}}} \, ,
	\label{}
\end{equation}
where $G$ reads as in eq.~\eqref{ex:Gappendix}, $\set{L^{\lambda_{\mathrm{b}}}}$ are the broken components and $\set{f_{\lambda_{\mathrm{b}}}}$ are here interpreted as the conjugated generalized forces.
Adding and subtracting the term $Z^{\sigma} \mu^{\mathrm{eq}}_{\sigma} + Z_{0} \mu^{\mathrm{eq}}_{0}$ from the last equation and using eq.~\eqref{eq:mueqSolvent} we obtain
\begin{equation}
	\mathcal{G} = \mathcal{G}_{\mathrm{eq}} + Z^{\sigma} \left( \mu_{\sigma} - \mu^{\mathrm{eq}}_{\sigma} \right) + Z^{0} \left( \mu_{0} - \mu^{\mathrm{eq}}_{0} \right) \, ,
	\label{}
\end{equation}
where
\begin{equation}
	\mathcal{G}_{\mathrm{eq}} = f_{\lambda_{\mathrm{u}}} \, L^{\lambda_{\mathrm{u}}} \, .
	\label{}
\end{equation}
From eqs.~\eqref{ex:cpSolutes} and \eqref{ex:cpSolvent} and the fact that $Z^{\sigma_{\mathrm{y}}} = Z_{\mathrm{eq}}^{\sigma_{\mathrm{y}}}$ and $Z_{0} = Z^{\mathrm{eq}}_{0}$ we obtain
\begin{equation}
	\begin{split}
		\mathcal{G} & \simeq \mathcal{G}_{\mathrm{eq}} + Z^{\sigma_{\mathrm{x}}} \left( \mu_{\sigma_{\mathrm{x}}} - \mu^{\mathrm{eq}}_{\sigma_{\mathrm{x}}} \right) - RT \left( Z^{\mathcal{S}_{\mathrm{x}}} - Z^{\mathcal{S}_{\mathrm{x}}}_{\mathrm{eq}} \right) \\
		& = \mathcal{G}_{\mathrm{eq}} + Z^{\sigma_{\mathrm{x}}} \, RT \ln \frac{Z^{\sigma_{\mathrm{x}}}}{Z_{\mathrm{eq}}^{\sigma_{\mathrm{x}}}} - RT \left( Z^{\mathcal{S}_{\mathrm{x}}} - Z^{\mathcal{S}_{\mathrm{x}}}_{\mathrm{eq}} \right) \\
		& \equiv \mathcal{G}_{\mathrm{eq}} + RT \, \mathcal{L} ( \{ Z^{\sigma_{\mathrm{x}}} \} | \{ Z_{\mathrm{eq}}^{\sigma_{\mathrm{x}}} \} )
	\end{split}
	\label{}
\end{equation}
in agreement with the expression derived in the main text, eq.~\eqref{eq:tGLtGeq}.

\section{Entropy of CRNs}
\label{sec:stochEntropy}

We show how the nonequilibrium entropy \eqref{ex:se} can be obtained as a large particle limit of the stochastic entropy.
We point out that while finalizing the paper similar derivations for other thermodynamic quantities have obtained in Refs.~\cite{ge16,ge16:mesoscopic}.

In the stochastic description of CRNs, the state is characterized by the population vector $\mathbf{n} = \set{n^{\sigma}}$.  
The probability to find the network is in state $\mathbf{n}$ at time $t$ is denoted $p_{t}(\mathbf{n})$.
The stochastic entropy of that state reads \cite{schmiedl07,esposito12}, up to constants,
\begin{equation}
	S(\mathbf{n}) = - \kb \ln p_{t}(\mathbf{n}) + s(\mathbf{n}) \, .
	\label{def:ses}
\end{equation}
The first term is a Shannon-like contribution while the second term is the \emph{configurational entropy}
\begin{equation}
	s(\mathbf{n}) \equiv n^{\sigma} {\tilde{s}^\circ_\sigma} - \kb \sum_{\sigma} \ln \frac{n^{\sigma}!}{n_{0}^{n^{\sigma}}} \, .
	\label{ex:ce}
\end{equation}
$\tilde{s}^\circ_{\sigma}$ is the standard entropy of one single $\ce{X}_\sigma$ molecule, and $n_{0}$ is the very large number of solvent molecules.

We now assume that the probability becomes very narrow in the large particle limit $n^{\sigma} \gg 1$ and behaves as a discrete delta function $p_{t}(\mathbf{n}) \simeq \delta(\mathbf{n} - \hat{\mathbf{n}} (t))$.
The vector $\hat{\mathbf{n}}(t) \equiv \set{\hat{n}^{\sigma}}$ denotes the most probable and macroscopic amount of chemical species, such that $Z^{\sigma} = {\hat{n}^{\sigma}}/(V N_{\mathrm{A}})$.
Hence, the average entropy becomes 
\begin{equation}
	\ave{S} = \sum_{\mathbf{n}} p_{t}(\mathbf{n}) S(\mathbf{n}) \simeq s(\hat{\mathbf{n}}) \, .
	\label{}
\end{equation}
When using the Stirling approximation ($\ln m! \simeq m \ln m - m$ for $m \gg 1$), we obtain
\begin{equation}
	\begin{split}
		s(\hat{\mathbf{n}}) & \simeq \hat{n}^{\sigma} \tilde{s}^\circ_{\sigma} - \hat{n}^{\sigma} \, \kb \ln \frac{\hat{n}^{\sigma}}{n_{0}} + \kb \sum_{\sigma} \hat{n}^{\sigma} \\
		& = \hat{n}^{\sigma} \left( \tilde{s}^\circ_{\sigma} + \kb \ln \frac{n_{0}}{V N_{\mathrm{A}}} \right) + \\
		& \quad - \hat{n}^{\sigma} \, \kb \ln \frac{\hat{n}^{\sigma}}{V N_{\mathrm{A}}} + \kb \sum_{\sigma} \hat{n}^{\sigma} \\
		& \equiv \hat{n}^{\sigma} \left( \tilde{s}^\circ_{\sigma} + \kb \ln Z_{0} \right) + \\
		& \quad - \hat{n}^{\sigma} \, \kb \ln Z^{\sigma} + \kb \sum_{\sigma} \hat{n}^{\sigma} \, .
	\end{split}
	\label{}
\end{equation}
Dividing by $V$ and using the relation $R=N_{\mathrm{A}} \kb$ we finally get the macroscopic entropy density \eqref{ex:se}
\begin{equation}
	{\ave{S}}/{V} \simeq Z^{\sigma} \, s^\circ_{\sigma} - Z^{\sigma} \, R \ln Z_{\sigma} + R Z^{\mathcal{S}} \, ,
	\label{}
\end{equation}
where the (molar) standard entropies of formation $s^\circ_{\sigma}$ reads
\begin{equation}
	s^\circ_{\sigma} = N_{\mathrm{A}} \left( \tilde{s}^\circ_{\sigma} + \kb \ln Z_{0} \right) \, .
	\label{}
\end{equation}

Mindful of the information-theoretical interpretation of the entropy \cite{jaynes57}, we note that the uncertainty due to the stochasticity of the state disappears (the first term on the r.h.s. of eq.~\eqref{def:ses}).
However, the uncertainty due to the indistinguishability of the molecules of the same species---quantified by the configurational entropy \eqref{ex:ce}---remains and contributes to the whole deterministic entropy function \eqref{ex:se}.

\section{Adiabatic--Nonadiabatic Decomposition}
\label{sec:anadec}

We prove the positivity of the adiabatic and nonadiabatic entropy production rates \eqref{ex:ANAdec} using the theory of complex-balanced networks, see \S~\ref{sec:cbn}.

We first rewrite the mass action kinetics currents \eqref{ex:currents} as \cite{gunawardena03,schaft15}:
$J^{\rho} = K^{\rho}_{\gamma'} \, \psi^{\gamma'}$, where $\psi^{\gamma} \equiv {Z^{\sigma}}^{\Gamma^{\gamma'}_{\sigma}}$ and ${K} = \set{K^{\rho}_{\gamma} \equiv K^{+\rho}_{\gamma} - K^{-\rho}_{\gamma}}$ is the \emph{rate constants matrix} whose entries are defined by
\begin{equation}
	K^{\rho}_{\gamma} = 
	\begin{cases}
		k^{+\rho} & \text{if $\gamma$ is the reactant complex of $+\rho$} \, , \\
		- k^{-\rho} & \text{if $\gamma$ is the product complex of $+\rho$} \, , \\
		0 & \text{otherwise} \, .
	\end{cases}
	\label{}
\end{equation}
Hence, the definition of complex-balanced network \eqref{def:cb} reads
\begin{equation}
	\sum_{\gamma \in \mathcal{C}_{j}} \mathcal{W}^{\gamma}_{\gamma'} \, \bar{\psi}^{\gamma'} = 0 \, , \quad \forall j \, .
	\label{eq:cbW}
\end{equation}
where $\mathcal{W} \equiv \partial \, {K} = \set{\partial^{\gamma}_{\rho} \, K^{\rho}_{\gamma'}} \equiv \set{\mathcal{W}^{\gamma}_{\gamma'}}$ is the so-called \emph{kinetic matrix} \cite{horn72}, and $\bar{\psi}^{\gamma} \equiv \left.{{\bar{Z}}^{\sigma}}\right.^{\Gamma^{\gamma}_{\sigma}}$.

The kinetic matrix $\mathcal{W} $ is a \emph{Laplacian matrix} \cite{schaft15,schaft13}:
any off-diagonal term is equal to the rate constant of the reaction having $\gamma'$ as a reactant and $\gamma$ as product if the reaction exists, and it is zero otherwise.
Also, it satisfies
\begin{equation}
	\sum_{\gamma \in \mathcal{C}} \mathcal{W}^{\gamma}_{\gamma'} = 0 \, ,
	\label{eq:sumcolumns}
\end{equation}
which is a consequence of the fact that the diagonal terms are equal to minus the sum of the off-diagonal terms along the columns.
The detailed balanced property \eqref{eq:db} implies that
\begin{equation}
	\mathcal{W}^{\gamma}_{\gamma'} \, \psi^{\mathrm{eq}}_{\gamma'} = \mathcal{W}^{\gamma'}_{\gamma} \, \psi^{\mathrm{eq}}_{\gamma} \, , \quad \forall \, \gamma , \gamma' \, ,
	\label{eq:dbW}
\end{equation}
where $\psi^{\mathrm{eq}}_{\gamma} \equiv \left.{Z^{\mathrm{eq}}_{\sigma}}\right.^{\Gamma_{\gamma}^{\sigma}}$.

In order to prove the non-negativity of the adiabatic term \eqref{ex:ANAdec}, we rewrite it as
\begin{equation}
	\begin{split}
		\vepr{a} & \equiv J^{\rho} \ln \frac{\bar{J}_{+\rho}}{\bar{J}_{-\rho}} 
		= K^{\rho}_{\gamma'} \, \psi^{\gamma'} \ln \left( \frac{\bar{Z}_{\sigma}}{{Z}^{\mathrm{eq}}_{\sigma}} \right)^{-\nabla^{\sigma}_{\rho}} \\
		& = - \mathcal{W}^{\gamma}_{\gamma'} \, \psi^{\gamma'} \ln \frac{\bar{\psi}_{\gamma}}{\psi^{\mathrm{eq}}_{\gamma}} \, .
	\end{split}
	\label{InterAd}
\end{equation}
The detailed balance property is used in the first equality, and the decomposition of the stoichiometric matrix \eqref{ex:NablaDec} in the second one.
Also, the constant $RT$ is taken equal to one.
Using \eqref{eq:sumcolumns}, eq.~\eqref{InterAd} can be rewritten as
\begin{equation}
	\vepr{a} = - \mathcal{W}^{\gamma}_{\gamma'} \, \psi^{\gamma'} \ln \frac{\bar{\psi}_{\gamma}\psi_{\mathrm{eq}}^{\gamma'}}{\psi^{\mathrm{eq}}_{\gamma} \bar{\psi}^{\gamma'}} \, .
	\label{}
\end{equation}
From the log inequality $-\ln x \ge 1 - x$ and the detailed balance property \eqref{eq:dbW}, we obtain
\begin{equation}
	\begin{split}
		\vepr{a} & \ge \mathcal{W}^{\gamma}_{\gamma'} \, \psi^{\gamma'} \left( 1 - \frac{\bar{\psi}_{\gamma}\psi_{\mathrm{eq}}^{\gamma'}}{\psi^{\mathrm{eq}}_{\gamma} \bar{\psi}^{\gamma'}} \right) \\
		& = - \mathcal{W}^{\gamma}_{\gamma'} \, \psi_{\mathrm{eq}}^{\gamma'} \, \frac{\bar{\psi}_{\gamma} \psi^{\gamma'}}{\psi^{\mathrm{eq}}_{\gamma} \bar{\psi}^{\gamma'}} = - \mathcal{W}^{\gamma'}_{\gamma} \, \bar{\psi}^{\gamma} \, \frac{\psi_{\gamma'}}{\bar{\psi}_{\gamma'}} = 0 \, .
	\end{split}
	\label{}
\end{equation}
The last equality follows from the assumption of complex-balanced steady state \eqref{eq:cbW}, the properties of the groups of complexes $\set{\mathcal{C}_{j}}$ (\S~\ref{sec:cbn}), and the fact that $\set{Z_{\sigma_{\mathrm{y}}} = \bar{Z}_{\sigma_{\mathrm{y}}}}$.
Indeed
\begin{equation}
	\hspace{-0.5cm}
	\begin{split}
		\mathcal{W}^{\gamma'}_{\gamma} \, \bar{\psi}^{\gamma} \, \frac{\psi_{\gamma'}}{\bar{\psi}_{\gamma'}} & =
		\sum_{j}^{} \sum_{\gamma' \in \mathcal{C}_{j}}^{} \mathcal{W}^{\gamma'}_{\gamma} \bar{\psi}^{\gamma} \left( \frac{{Z}_{\sigma_{\mathrm{x}}}}{\bar{Z}_{\sigma_{\mathrm{x}}}} \right)^{\Gamma^{\sigma_{\mathrm{x}}}_{\gamma'}} \\
		& = \sum_{j}^{} \left( \frac{{Z}_{\sigma_{\mathrm{x}}}}{\bar{Z}_{\sigma_{\mathrm{x}}}} \right)^{\Gamma^{\sigma_{\mathrm{x}}}_{j}} \sum_{\gamma' \in \mathcal{C}_{j}}^{} \mathcal{W}^{\gamma'}_{\gamma} \bar{\psi}^{\gamma} = 0 \, .
	\end{split}
	\label{eq:proof}
\end{equation}

Concerning the nonadiabatic term \eqref{ex:ANAdec}, using the rate equations \eqref{ex:intflux} and the fact that $\set{Z_{\sigma_{\mathrm{y}}} = \bar{Z}_{\sigma_{\mathrm{y}}}}$, we can rewrite it as
\begin{equation}
		\vepr{na} \equiv - \dt{Z}^{\sigma} \ln \frac{Z_{\sigma}}{\bar{Z}_{\sigma}} 
			= - \mathcal{W}^{\gamma}_{\gamma'} \, \psi^{\gamma'} \ln \frac{\psi_{\gamma}}{\bar{\psi}_{\gamma}} \, .
	\label{}
\end{equation}
Because of \eqref{eq:sumcolumns}, we further get that
\begin{equation}
	\vepr{na} = - \mathcal{W}^{\gamma}_{\gamma'} \, \psi^{\gamma'} \ln \frac{\psi_{\gamma}\bar{\psi}^{\gamma'}}{\bar{\psi}_{\gamma}\psi^{\gamma'}} \, .
	\label{}
\end{equation}
From the log inequality $-\ln x \ge 1 - x$ and from \eqref{eq:dbW} 
\begin{equation}
\hspace{-0.5cm} \vepr{na} \ge \mathcal{W}^{\gamma}_{\gamma'} \, \psi^{\gamma'} \left( 1 - \frac{\psi_{\gamma}\bar{\psi}^{\gamma'}}{\bar{\psi}_{\gamma}\psi^{\gamma'}} \right)
= - \mathcal{W}^{\gamma}_{\gamma'} \, \bar{\psi}^{\gamma'} \, \frac{\psi_{\gamma}}{\bar{\psi}_{\gamma}} = 0 \, . \label{}
\end{equation}
The last equality again follows from the assumption of complex-balance steady state \eqref{eq:cbW} as in eq.~\eqref{eq:proof}.

\section{Deficiency of CRNs}
\label{sec:dzn}

The deficiency of an open CRN is defined as \cite{polettini15}
\begin{equation}
	\delta = \dim \ker \nabla^{\mathrm{X}} - \dim \ker \partial^{\mathcal{C}} \geq 0 \, ,
	\label{def:deficiency}
\end{equation}
where $\partial^{\mathcal{C}} = \set{{\partial}^{j}_{\rho} \equiv \sum_{\gamma \in \mathcal{C}_{j}} \partial^{\gamma}_{\rho}}$.
Other equivalent definitions can be found in \cite{feinberg:lectures} and \cite{gunawardena03}.
The kernel of $\nabla^{\mathrm{X}}$ identifies the set of cycles, eqs.~\eqref{def:cycle} and \eqref{def:emergent}, while the kernel of the incidence matrix $\hat{\partial}$ identifies the set of cycles of the reaction graph. 
Hence, the deficiency measures the difference between the number of cyclic transformations on chemical species and how many of them can be represented as cycles on the reaction graph. Deficiency-zero networks are defined by $\delta = 0$, \emph{i.e.} they exhibit a one to one correspondence between the two.
This topological property has many dynamical consequences, the most important of which is that deficiency-zero networks are unconditionally complex-balanced \cite{horn72:complex,feinberg72}.
As shown in Ref.~\cite{polettini15}, deficiency has also implications on the stochastic thermodynamic description of networks:
the stochastic entropy production of a deficiency-zero network converges to the deterministic entropy production in the long-time limit.
Linear networks are the simplest class of deficiency-zero networks.
Since only one internal species appears in each complex with a stoichiometric coefficient equal to one,
$\nabla^{\mathrm{X}} \equiv \partial^{\mathcal{C}}$ and thus $\delta = 0$.

\bibliography{indiceLocale}

\end{document}